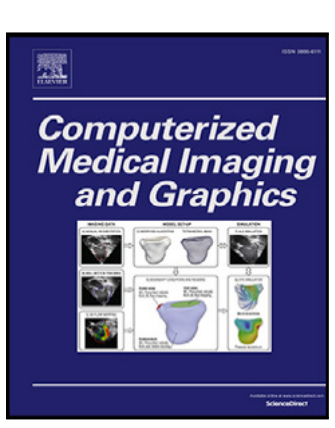

# UltraBoneUDF: Self-supervised bone surface reconstruction from ultrasound based on neural unsigned distance functions

Luohong Wu [a,*], Matthias Seibold [a], Nicola A. Cavalcanti [a], Giuseppe Loggia [b], Lisa Reissner [b], Bastian Sigrist [a], Jonas Hein [a,c], Lilian Calvet [a], Arnd Viehöfer [b], Philipp Fürnstahl [a]

[a] Research in Orthopedic Computer Science, Balgrist University Hospital, University of Zurich, Lengghalde 5, Zurich, 8008, Switzerland
[b] Department of Orthopaedics, Balgrist University Hospital, University of Zurich, Forchstrasse 340, Zurich, 8008, Switzerland
[c] Computer Vision and Geometry Group, ETH Zurich, Ramistrasse 101, Zurich, 8092, Switzerland

ABSTRACT

**Background:** Bone surface reconstruction is an essential component of computer-assisted orthopedic surgery (CAOS), forming the foundation for both preoperative planning and intraoperative guidance. Compared to traditional imaging modalities such as computed tomography (CT) and magnetic resonance imaging (MRI), ultrasound, an emerging CAOS technology, provides a radiation-free, cost-effective, and portable alternative. While ultrasound offers new opportunities in CAOS, technical shortcomings continue to hinder its translation into surgery. In particular, due to the inherent limitations of ultrasound imaging, B-mode ultrasound typically captures only partial bone surfaces. The inter- and intra-operator variability in ultrasound scanning further increases the complexity of the data. Existing reconstruction methods struggle with such challenging data, leading to increased reconstruction errors and artifacts, such as holes and inflated structures. Effective techniques for accurately reconstructing open bone surfaces from real-world 3D ultrasound volumes remain lacking.
**Methods:** We propose UltraBoneUDF, a self-supervised framework specifically designed for reconstructing open bone surfaces from ultrasound data. It learns unsigned distance functions (UDFs) from 3D ultrasound data. In addition, we present a novel loss function based on local tangent plane optimization that substantially improves surface reconstruction quality. UltraBoneUDF and competing models are benchmarked on three open-source datasets and further evaluated through ablation studies.
**Results:** Qualitative results demonstrate the limitations of the state-of-the-art methods. Quantitatively, UltraBoneUDF achieves comparable or lower bi-directional Chamfer distance across three datasets with fewer parameters: 1.60 mm on the UltraBones100k dataset ($\approx 25.5\%$ improvement), 0.21 mm on the OpenBoneCT dataset, and 0.18 mm on the ClosedBoneCT dataset.
**Conclusion:** UltraBoneUDF represents a promising solution for open bone surface reconstruction from 3D ultrasound volumes, with the potential to advance downstream applications in CAOS.

## 1. Introduction

Computer-Assisted Orthopedic Surgery (CAOS) has demonstrated significant potential to enhance surgical precision in a range of procedures (Mavrogenis et al., 2013), including bone tumor resection (Evrard et al., 2022), total knee arthroplasty (Sugano, 2003), and deformity correction (Baraza et al., 2020). Central to CAOS is the availability of precise 3D representations of target bones, which enable effective preoperative planning. These representations also support intraoperative visualization (Fadero and Shah, 2014; Tetsworth et al., 2017) and guidance through navigation systems (So et al., 2010; Hohlmann et al., 2024). Current workflows predominantly rely on diagnostic imaging modalities such as Computed Tomography (CT) and Magnetic Resonance Imaging (MRI). While these modalities provide high-resolution anatomical detail, their intraoperative use is limited by radiation exposure, high cost, and restricted intraoperative applicability.

Ultrasound imaging represents a compelling complementary modality, providing radiation-free, portable, and real-time acquisition capabilities. Recent advances in 3D reconstruction from ultrasound data demonstrate that intraoperative ultrasound can provide volumetric anatomical information. This extends beyond conventional two-dimensional visualization. A growing body of literature demonstrates its potential across various CAOS scenarios. For example, in the preoperative stage, ultrasound-based bone reconstruction supports implant design and surgical planning in procedures such as total knee arthroplasty (TKA) (Hohlmann et al., 2024; Hohlmann, 2024) and total hip arthroplasty (THA) (Gebhardt et al., 2023; Schumann et al., 2012; Guezou-Philippe et al., 2023), eliminating radiation exposure inherent to conventional CT-based 3D reconstruction. In the intraoperative stage, it facilitates CT/MRI–ultrasound registration for transferring preoperative plans (Ciganovic et al., 2018; van der Zee et al., 2023). However, most existing work remains at the proof-of-concept or preclinical stage, as reliable and precise ultrasound segmentation and 3D reconstruction continue to face limitations that hinder reliable implementation into surgical practice.

* Correspondence to: Lengghalde 5, Zurich, 8008, Switzerland
*E-mail address:* luohong.wu@balgrist.ch (L. Wu).

2D B-mode ultrasound forms images from reflected acoustic waves (Wu et al., 2025). Because bone has the highest acoustic impedance among soft tissues, superficial bone surfaces reflect much of the incident energy, producing hyperechoic (bright) bands. An anechoic (dark) region typically appears beneath, known as bone shadowing (Pandey et al., 2020). Shadowing limits ultrasound visibility to only partial bone surfaces, resulting in topologically open bone surfaces being captured. For example, in in-vivo and ex-vivo scans of the spine, only the posterior surface of the vertebrae is visible, whereas the anterior surface is completely missing due to acoustic shadowing (Li et al., 2023). In general, comprehensive imaging of bony anatomy from all sides is often impractical due to probe placement constraints, overlying structures, and limited acoustic access. Consequently, clinical acquisitions commonly contain only partial bone surfaces, underscoring the need for methods that reconstruct high-quality open bone surfaces from 3D ultrasound volumes.

A common approach for reconstructing 3D bone models from an ultrasound sweep involves compounding 2D ultrasound images spatially by tracking the ultrasound probe (Mozaffari and Lee, 2017). Subsequently, 2D ultrasound segmentation is utilized to localize the bone anatomy in each slice and reconstruct the 3D model based on the tracking data (Hohlmann et al., 2024). Previous studies have introduced methods for reconstructing continuous bone surfaces from sparse point clouds (Mozaffari and Lee, 2017), including Poisson surface reconstruction (Gebhardt et al., 2023) and statistical shape modeling (SSM) (Hohlmann et al., 2023; Mahfouz et al., 2021). However, even with computationally expensive post-processing steps such as smoothing and interpolation, the resulting mesh quality remains inadequate. Visible artifacts, reduced accuracy, and limited robustness to input noise persist (Nguyen et al., 2015).

These limitations have motivated a shift toward alternative representations. In this context, Implicit Neural Representations (INRs) have gained popularity in computer vision research (Mildenhall et al., 2021) and have also been applied to learn Signed Distance Functions (SDFs) from freehand ultrasound volumes, enabling the accurate reconstruction of closed anatomical surfaces (Chen et al., 2024b,a). These methods learn to partition the target space into interior and exterior regions, with the anatomical surface reconstructed as the boundary between these two regions (Chen et al., 2024b,a). SDF-based methods outperform traditional approaches in terms of robustness and accuracy, particularly for closed objects. However, it is well-known in the computer vision state-of-the-art (SOTA) that SDFs are inadequate for reconstructing topologically open surfaces, whereas Unsigned Distance Functions (UDFs) are better suited for such data (Guillard et al., 2022; Zhou et al., 2024). SOTA UDF-based methods typically assume access to high-quality point clouds for learning UDFs. These point clouds are usually synthetic or collected in a controlled setting to ensure continuity and density. The properties enable accurate closest-distance querying (Chibane and Pons-Moll, 2020), as well as continuous and smooth gradient sign flips (Zhou et al., 2024). Their effectiveness in reconstructing high-fidelity open surfaces has been demonstrated in previous studies (Chibane and Pons-Moll, 2020; Venkatesh et al., 2020; Atzmon and Lipman, 2020; Guillard et al., 2022; Zhou et al., 2024).

However, transferring these methods to real-world ultrasound data is not straightforward because of domain-specific challenges. Real-world 3D freehand ultrasound data are manually acquired by clinical experts along a specific direction (e.g., longitudinal for the fibula), including both inter- and intra-operator variability. The single scanning direction leads to slice-wise structures in the point clouds. Variations in probe speed and incidence angle between the probe and target bone introduce spatial gaps (Wu et al., 2025), while the synthetic data exhibit a more continuous and uniform distribution. Furthermore, segmentation and tracking errors in real clinical data make the point clouds noisier and more challenging to reconstruct. Although existing UDF methods can reconstruct bone meshes from such data, adapting them to account for domain-specific challenges requires additional research and methodologies. Nonetheless, this remains largely unexplored.

In this work, we bridge this gap by providing methodologies and adaptions, as well as benchmarking various existing design choices. We propose UltraBoneUDF, a self-supervised model for learning UDF to accurately reconstruct open bone surfaces from freehand 3D ultrasound volumes. It is comprehensively evaluated against existing methods on three publicly available datasets.

The main contributions include:

- Inspired by prior work (Zhou et al., 2024; Chibane and Pons-Moll, 2020), UltraBoneUDF directly learns unsigned distance functions from ultrasound point clouds, avoiding sign ambiguities and inflated artifacts. The results demonstrate that UltraBoneUDF matches or surpasses the performance of SOTA methods while using fewer parameters and requiring less computation time.
- We propose a tangent-plane–based loss with dynamic neighbor updates to stabilize training. This loss offers improved initialization for downstream tangent-plane–based optimization in mesh extraction methods, such as DualMesh-UDF (Zhang et al., 2023). Ablation studies show that our loss achieves lower surface distance errors.
- We present qualitative and quantitative evaluations and analyze the limitations of the current SOTA SDF-based method for reconstructing open bone surfaces from 3D freehand ultrasound. We also conduct experiments on three open-source datasets to comprehensively evaluate and benchmark different design choices.

To ensure reproducibility of the results, the code has been made publicly available.[1]

## 2. Related work

**3D Bone Reconstruction from Ultrasound**. Traditionally, 3D geometry from ultrasound is obtained by tracking the probe during acquisition and projecting the corresponding 2D segmentation masks into 3D space to form point clouds (Hohlmann et al., 2024; Ciganovic et al., 2018; van der Zee et al., 2023). Continuous bone surfaces can be reconstructed from point clouds using Poisson reconstruction (Gebhardt et al., 2023) or statistical shape modeling (Hohlmann et al., 2023; Mahfouz et al., 2021). However, these approaches often require complex, input-dependent post-processing steps, such as smoothing and interpolation, to enhance reconstruction quality (Berger et al., 2017). They also struggle with the reconstruction of complex surface geometries, which result in artifacts and increased surface errors. Moreover, these methods are not robust to input noise, such as tracking errors and segmentation inaccuracies (Nguyen et al., 2015). Recently, deep learning models have been developed to reconstruct bone meshes directly from sparse ultrasound point clouds (Ma et al., 2020). Chen et al. introduced FUNSR, an online self-supervised neural implicit surface reconstruction model, to reconstruct femur and pelvis meshes from bone segmentations of tracked ultrasound images (Chen et al.,

[1] https://github.com/luohwu/UltraBoneUDF

2024b). In a first step, the method converts tracked 2D ultrasound bone segmentation masks into a volumetric point cloud. Subsequently, FUNSR trains a Multi-Layer Perceptron (MLP)-based neural network to learn a SDF that describes the surface of the bone anatomy. The shape space is then partitioned into inside and outside regions. Finally, the Marching Cubes algorithm is applied to extract high-quality meshes from the learned SDFs (Lorensen and Cline, 1998). Building on FUNSR, RoCoSDF merges SDFs learned from multiple ultrasound scans of the same anatomy to learn a more precise SDF, thereby enhancing reconstruction accuracy (Chen et al., 2024a). However, SDF-based methods encounter significant challenges when reconstructing thin and open surfaces with complex topologies, as the shape space cannot be distinctly partitioned into interior and exterior regions (Venkatesh et al., 2020; Chibane and Pons-Moll, 2020). Consequently, the reconstructed meshes exhibit artifacts such as holes or inflated structures, leading to increased reconstruction errors.

**Unsigned Distance Functions-based Surface Reconstruction**. Instead of partitioning the shape space into interior and exterior regions and computing signed distances, UDFs directly calculate the absolute distance of a point to the target surface. Julian et al. introduced NDF, a supervised neural model that learns an unsigned distance function directly from point clouds. NDF regresses the closest-point distance and enforces distance-like gradient behavior, enabling reconstruction of both topologically open and closed objects (Chibane and Pons-Moll, 2020). Similarly, Rahul et al. developed DUDE, a method for learning deep unsigned distance embeddings to achieve high-fidelity representations of complex 3D surfaces by supervising both distance values and normal vectors (Venkatesh et al., 2020). In contrast to neural SDFs, the Marching Cubes algorithm cannot be directly applied to neural UDFs, as the absence of sign transitions and the instability of the zero level set prevent reliable surface reconstruction. To address this limitation, computationally expensive methods like ray tracing can be employed (Chibane and Pons-Moll, 2020; Venkatesh et al., 2020). Matan et al. introduced SAL, a technique for generating SDFs from unsigned geometric data (Atzmon and Lipman, 2020). While the Marching Cubes algorithm can then be directly applied, the reconstructed surfaces frequently exhibit incorrectly closed meshes (Venkatesh et al., 2020). To efficiently extract meshes from learned UDFs, methods such as MeshUDF (Guillard et al., 2022) and CAP-UDF (Zhou et al., 2024) infer gradients of UDFs and apply the Marching Cubes algorithm based on the sign flips of gradients. These methods are primarily evaluated on synthetic datasets such as ShapeNet (Chang et al., 2015) and MGN (Bhatnagar et al., 2019), which allow uniform and continuous point sampling. Some works also test on real-world scanned datasets like KITTI (Zhou et al., 2024), where dense and continuous point clouds are available as well. The continuity and density of such data enable accurate closest-distance querying (Chibane and Pons-Moll, 2020), as well as continuous and smooth gradient sign flips (Zhou et al., 2024). However, real-world CAOS ultrasound data exhibit slice-wise structures, spatial gaps, segmentation and tracking errors, suggesting that further improvements are possible when these challenges are explicitly considered.

**UDF-based Surface Reconstruction from Ultrasound**. Chen et al. proposed a method for learning a neural UDF from 3D ultrasound data for closed carotid surface reconstruction (Chen et al., 2023). To the best of our knowledge, this is the only neural UDF-based approach for surface reconstruction from 3D ultrasound data. For simplicity, we refer to this method as CarotidUDF in the following sections. We will show in Section 3.3.2 that the loss function used by CarotidUDF can result in locally distorted unsigned distance field, leading to increased surface distance errors. More importantly, CarotidUDF converts the learned UDF into an SDF based on gradients for surface reconstruction. When applied to open bone surfaces, this process produces dual-layered or inflated structures (Chen et al., 2023), leading to higher surface distance error. This renders CarotidUDF ineffective for thin and open bone surface reconstruction. In contrast, our method, UltraBoneUDF, is specifically designed for reconstructing open bone surfaces from 3D ultrasound data. It takes into account the characteristics of open bone ultrasound data and the requirements of actual surgical scenarios, addressing the limitations of CarotidUDF and FUNSR for thin and open bone surface reconstruction.

## 3. Method

### 3.1. Problem formulation: 3D surface reconstruction with distance fields

We follow the preprocessing pipeline commonly used in SOTA for processing 3D ultrasound data (Chen et al., 2024b,a). Specifically, by using the calibration and tracking data of the ultrasound probe, pixels of the segmentation masks can be transformed into a 3D point cloud $\mathcal{P}_{\text{raw}}$ representing the underlying bone surface $\mathcal{S}$ (Chen et al., 2024b). This point cloud $\mathcal{P}_{\text{raw}}$ is subsequently spatially discretized into voxel grids and downsampled to achieve a uniform density (Chen et al., 2024b), denoted as $\mathcal{P} = \left\{ \mathbf{p}_i = \left[x_i, y_i, z_i\right]^T \mid i \in [1, N] \right\}$. Around each point $\mathbf{p}_i \in \mathcal{P}$, $M$ query points are sampled following a Gaussian distribution $\mathcal{N}(0, \delta)$, forming the query point cloud $\mathcal{Q} = \left\{ \mathbf{q}_{ij} \mid i \in [1, N], j \in [1, M] \right\}$ where $\delta$ is defined as the Euclidean distance between $\mathbf{p}_i$ and its $k$th nearest neighbor (Ma et al., 2020). For simplicity, we denote $L = M \cdot N$ and redefine $\mathcal{Q} = \left\{ \mathbf{q}_i \mid i \in [1, L] \right\}$, $\mathcal{T} = \left\{ \mathbf{t}_i \mid i \in [1, L] \right\}$, with

$$\mathbf{t}_i = f(\mathbf{q}_i) = \arg\min_{\mathbf{p} \in \mathcal{P}} \|\mathbf{q}_i - \mathbf{p}\| \tag{1}$$

Note that $\mathcal{T}$ contains duplicate points, as it represents the set of nearest points from $\mathbf{p}_i \in \mathcal{P}$ to each query point $\mathbf{q}_i \in \mathcal{Q}$. The goal is to reconstruct the underlying bone surface $\mathcal{S}$ from $\mathcal{Q}$ and $\mathcal{T}$.

### 3.2. Performance analysis of SDF-based bone surface reconstruction

The SOTA for reconstructing bone surfaces like FUNSR learn SDFs directly from raw ultrasound point clouds to reconstruct closed anatomical surfaces (Chen et al., 2024b). Specifically, an MLP is trained to represent the SDF $s : \mathbb{R}^3 \to \mathbb{R}$. A point $\mathbf{q}_i$ can be projected to $\mathbf{q}'_i$ near the predicted target surface using the equation:

$$\mathbf{q}'_i = \mathbf{q}_i - s(\mathbf{q}_i) \times \frac{\nabla s(\mathbf{q}_i)}{\|\nabla s(\mathbf{q}_i)\|_2} \tag{2}$$

, where $s(\mathbf{q}_i)$ is the predicted SDF value at $\mathbf{q}_i$ and $\frac{\nabla s(\mathbf{q}_i)}{\|\nabla s(\mathbf{q}_i)\|_2}$ represents the normalized gradient. The primary self-supervised loss function is formulated as:

$$\mathcal{L}_{\text{pull}} = \frac{1}{K} \sum_{k \in [1,K]} \|\mathbf{q}'_k - \mathbf{t}_k\|_2^2 \tag{3}$$

, where $K$ is the batch size. $\mathcal{L}_{\text{pull}}$ helps the network training process by pulling $\mathbf{q}'_k$ toward its static nearest neighbor $\mathbf{t}_k$. Additional loss terms are incorporated to enhance the reconstructed surface, including a sign consistency constraint and an on-surface constraint (Chen et al., 2024b). Previous work has proven that minimizing $\mathcal{L}_{\text{pull}}$ facilitates the network's convergence to the underlying SDF (Ma et al., 2020). However, the shape space of topologically open objects cannot be uniquely partitioned into interior and exterior regions. When SDF-based methods are applied to point clouds sampled from such objects, incorrect sign assignments often occur, particularly near surface boundaries. These sign errors can result in reconstruction artifacts such as holes or inflated structures. To illustrate this limitation, we applied FUNSR to a distorted closed sphere and an open semi-sphere. As shown in Fig. 1, the learned SDF for the semi-sphere exhibits noise around the surface boundary. Sign errors can not only introduce discontinuities and inflated surfaces, but also result in thicker surface reconstructions. In contrast, the SDF learned from the closed sphere remains consistent and artifact-free.

To provide an illustrative toy example of how surface characteristics affect reconstruction, we evaluate FUNSR on data derived from CT bone models with closed and open surfaces. As shown in Fig. 2, the model successfully reconstructs the closed bone surface but fails to accurately reconstruct the open surface.

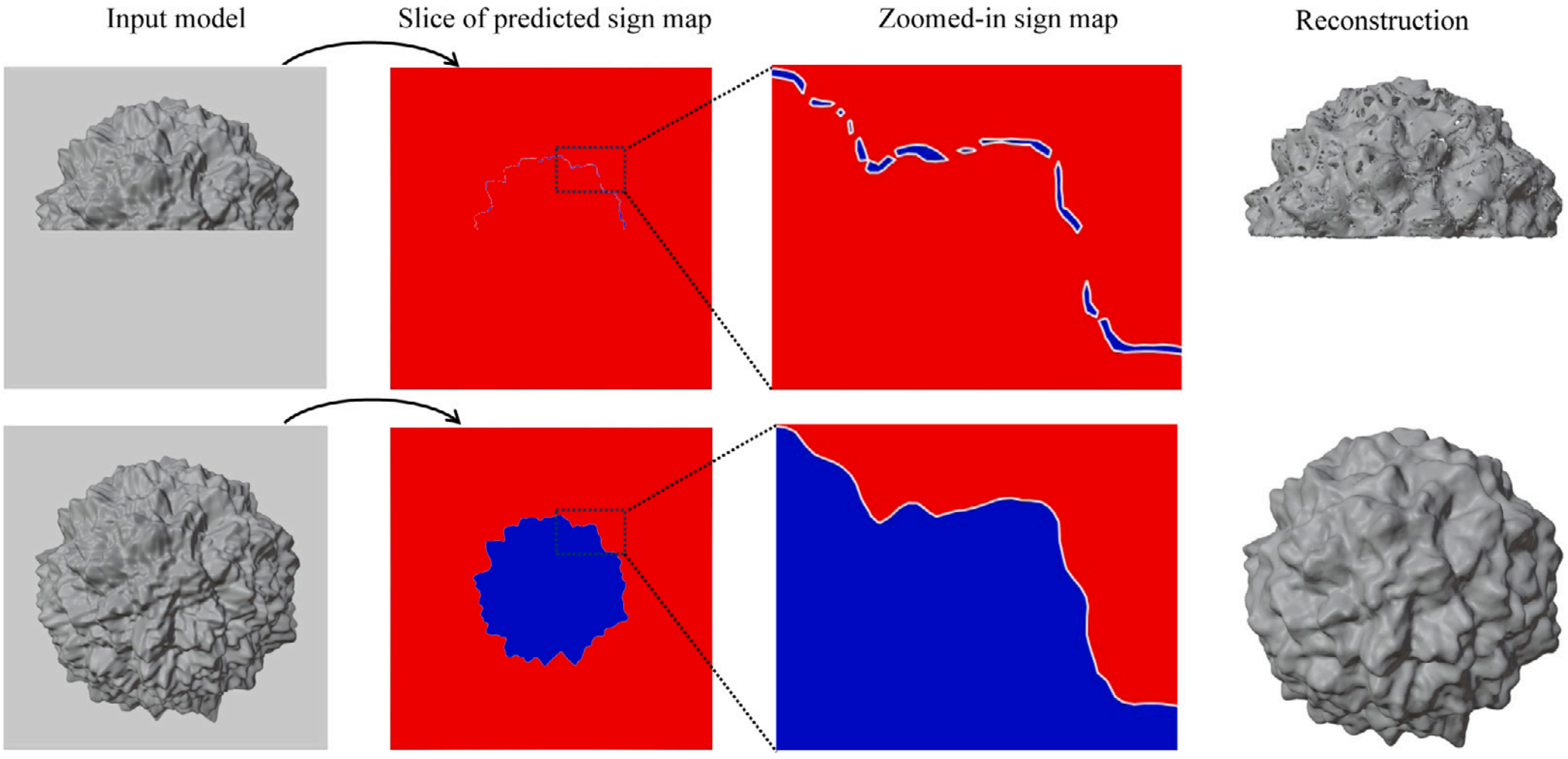

**Fig. 1.** Comparison of neural SDF reconstruction performance on a semi-sphere and a sphere, used here as illustrative synthetic example to demonstrate SDF limitations. First column: Input models used to generate sampled point clouds. Second column: Visualization of the learned sign assignment on the gray slicing plane depicted in the first column. The sign boundary is shown in white. Third column: Zoomed-in view of the sign map near the surface boundaries. Fourth column: Reconstructed 3D objects.

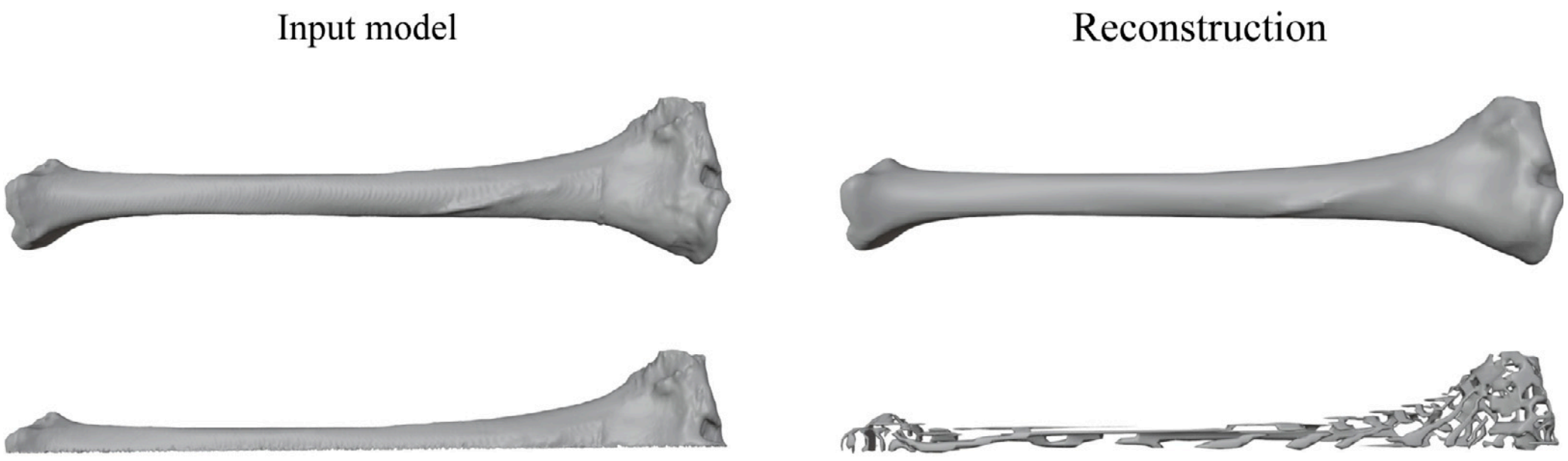

**Fig. 2.** Comparison of neural SDF reconstruction performance on a closed and open CT-derived bone model. The first column presents the GT CT-derived bone model, from which points are sampled to train the SDF-based method. The second column displays the corresponding reconstruction results.

To further explore this effect, we apply the solidify operation in Blender (Blender Institute, Amsterdam, Netherlands) to incrementally increase the thickness of the open bone model and apply FUNSR separately on each modified version. As illustrated in Fig. 3, increasing the thickness mitigates the hole issue. However, this also results in thicker reconstructions, leading to increased surface distance errors.

### 3.3. Our method

To address the limitations of the SDF-based methods, we introduce UltraBoneUDF, a model that learns neural UDFs from point clouds representing open and thin objects. UltraBoneUDF is a compact MLP with its architecture illustrated in Fig. 4. Given point coordinates $[x, y, z]$, UltraBoneUDF estimates the unsigned distance of query points $Q$ from the target surface. Similar to FUNSR, UltraBoneUDF is trained online in a self-supervised, per-instance manner, without requiring pretraining on other cases. Our loss is designed to address the error characteristics of neural UDFs, as detailed in Section 3.3.2, further enhancing reconstruction accuracy. Lastly, instead of converting the learned UDF to an SDF for surface reconstruction, we propose to employ DualMesh-UDF to extract the open bone surface directly from the learned neural implicit representation (Zhang et al., 2023), avoiding additional artifacts and reducing surface error.

#### 3.3.1. MLP-based UDF predictor

SDF-based approaches often rely on multiple sub-networks to enforce sign constraints and maintain correct sign predictions near surface boundaries, thereby increasing computational complexity (Ma et al., 2020; Chen et al., 2024b,a). UltraBoneUDF eliminates the need for sign prediction by employing a UDF predictor to directly learn the UDF. As shown in Fig. 4, the network consists of stacked fully connected layers followed by ReLU activations. The UDF predictor progressively maps 3D points in $\mathbb{R}^3$ to latent features of dimensions 64, 128, and 256, and finally outputs non-negative real values $u : \mathbb{R}^3 \rightarrow \mathbb{R}_{\geq 0}$, where each input $[x, y, z]$ denotes a 3D point and the output represents its normalized distance to the target surface. A normalization layer restores the normalized scale to its original range. Its gradient $\nabla u$ can be computed by a backpropagation operation. Similar to Eq. (2) in SDF-based methods, a query point $\mathbf{q}_i$ can be projected onto the surface learned by UltraBoneUDF as follows:

$$\mathbf{q}_i' = \mathbf{q}_i - u(\mathbf{q}_i) \times \frac{\nabla u(\mathbf{q}_i)}{\|\nabla u(\mathbf{q}_i)\|_2} \tag{4}$$

While UltraBoneUDF eliminates sign errors inherent to SDF-based methods, it is more challenging to train efficiently. To address this limitation, we propose a novel loss function in the following section.

#### 3.3.2. Loss function

**Local tangent plane optimization.** As demonstrated in prior work (Chen et al., 2024b; Zhou et al., 2024), both $\mathcal{T}$ and the mapping from $\mathcal{Q} \rightarrow \mathcal{T}$ are susceptible to imperfections and noise, due to factors such as sampling strategies, segmentation errors, and tracking inaccuracies. Direct optimization of $\mathcal{L}_{\text{pull}}$ in SDFs, such as FUNSR, leads to phenomena like sign inconsistency or oscillation, wherein the predicted sign for a given query point fluctuates during network training, particularly near surface boundaries. In UDFs, this manifests as oscillating optimization directions. A 2D example of this issue is shown in Fig. 5. For a detailed discussion of this problem in 3D, we refer to Zhou et al. (2024). To mitigate this problem, CAP-UDF proposed a Chamfer Distance (CD)-based loss function using dynamic nearest-neighbor querying. DualMesh-UDF further improves accuracy during the mesh extraction stage by optimizing a pretrained UDF with a tangent-plane–based objective that accounts for error characteristics of neural UDFs, such as approximation errors and spatial gaps (Zhang et al., 2023). Specifically, given a neural UDF encoding the target surface $\mathcal{S}$, DualMesh-UDF estimates the tangent planes of $\mathcal{S}$ at a set of sample points near $\mathcal{S}$. These points are grouped into local clusters. For each cluster, a linear least squares problem is solved to determine the final surface points, which are connected to construct the output surface (Zhang et al., 2023).

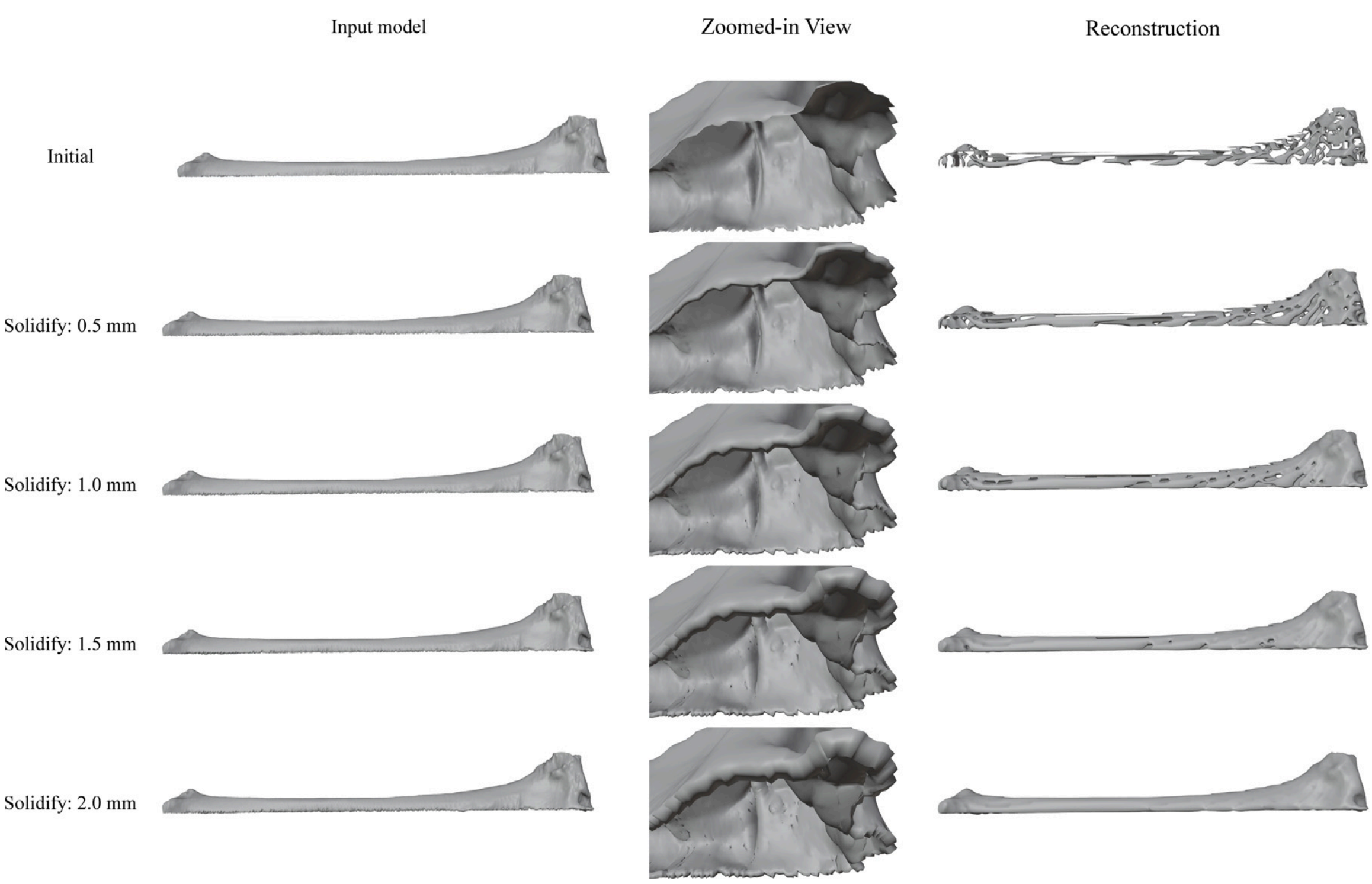

**Fig. 3.** Comparison of neural SDF reconstruction performance on open CT-derived bone models with varying thicknesses. The first row shows the reconstruction result on the initial open CT model. The subsequent rows present the reconstruction results on inflated CT-derived bone models, solidified with thicknesses of 0.5 mm, 1.0 mm, 1.5 mm, and 2.0 mm, respectively. The second column provides a zoomed-in view of the thickness for each corresponding CT-derived bone model.

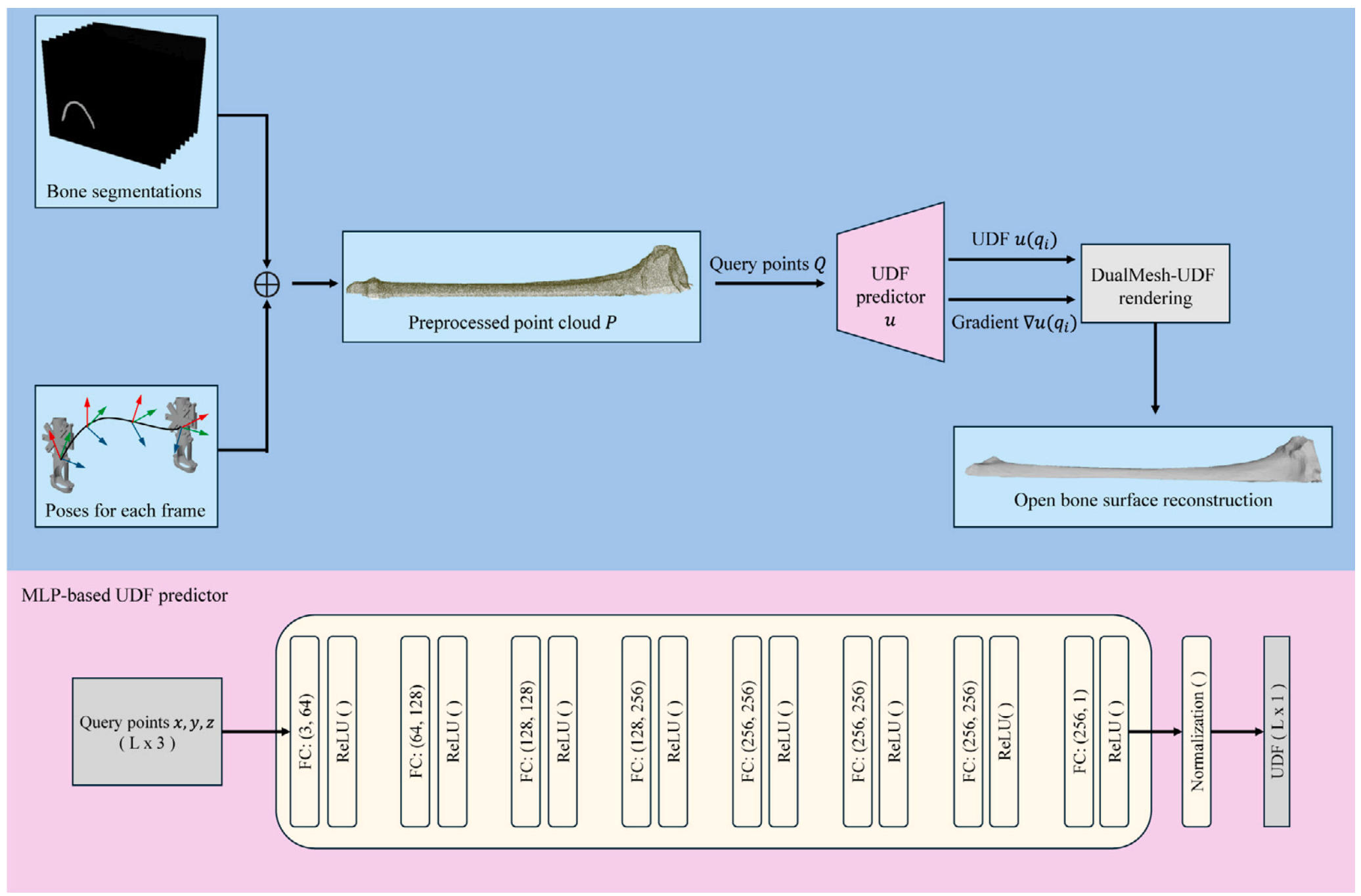

**Fig. 4.** Overview of the proposed method. Given bone segmentation masks and corresponding 3D poses, UltraBoneUDF learns the underlying UDFs, enabling precise and artifact-free target surface reconstruction.

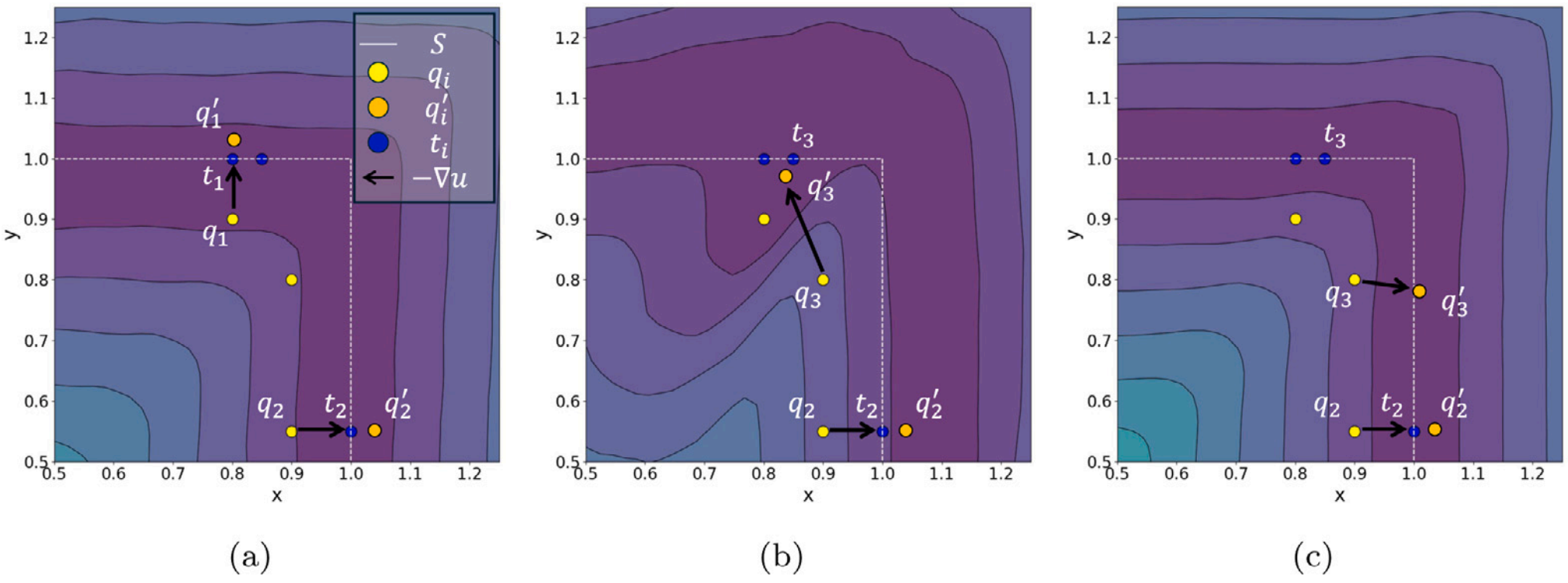

**Fig. 5.** Illustration of the oscillation problem in a 2D example when training with $\mathcal{L}_{\text{pull}}$. (a) Initial state at the $i$th iteration trained on $\mathbf{q}_1, \mathbf{q}_2$. (b) Result at the $(i+1)$-th iteration when trained on $\mathbf{q}_2, \mathbf{q}_3$ using $\mathcal{L}$pull, yielding an erroneous prediction. Depending on the sampled points in each iteration, the results can oscillate between (a) and (b). (c) Result at the $(i+1)$-th iteration when trained on $\mathbf{q}_2, \mathbf{q}_3$ using $\mathcal{L}_{\text{tangent}}$.

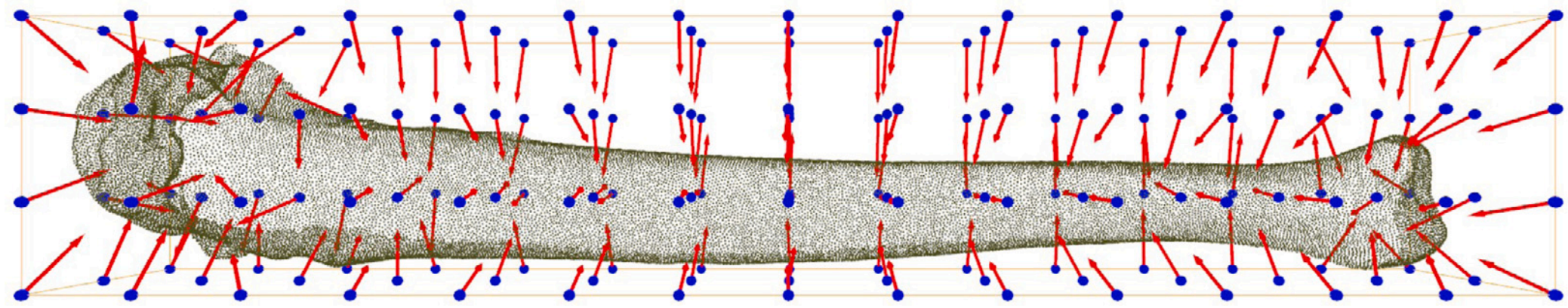

**Fig. 6.** Illustration of uniformly sampled grid points. Blue spheres represent the grid points, which are uniformly sampled within the bounding box of the input point cloud $\mathcal{P}$. Red arrows indicate the directional vectors from each sampled grid point to its nearest neighbor in $\mathcal{P}$.

Inspired by CAP-UDF (Zhou et al., 2024) and DualMesh-UDF (Zhang et al., 2023), we propose a loss function that implements dynamic nearest neighbor querying and tangent plane optimization. We hypothesize that a neural UDF trained with our loss can provide better initialization for the optimization during mesh extraction. Specifically, instead of relying on the static nearest neighbor mapping defined by Eq. (1), we dynamically update the nearest neighbor for a query point $\mathbf{q}_i$ during each batch iteration based on its projected point $\mathbf{q}_i'$ as follows:

$$f(\mathbf{q}_i') = \arg\min_{\mathbf{p}\in\mathcal{P}} \|\mathbf{q}_i' - \mathbf{p}\| \tag{5}$$

For instance, in Fig. 5(c), the projected nearest neighbor for $\mathbf{q}_3$ is $\mathbf{t}_2$ instead of $\mathbf{t}_3$. Furthermore, instead of pulling $\mathbf{q}_3$ directly toward $\mathbf{t}_2$ using $\mathcal{L}_{\text{pull}}$ as done in previous methods (Chen et al., 2024b,a; Ma et al., 2020), we propose the following loss function based on tangent plane optimization:

$$\mathcal{L}_{\text{tangent}} = \frac{1}{K} \sum_{i\in[1,K]} \left[\frac{\nabla u(\mathbf{q}_i)}{\|\nabla u(\mathbf{q}_i)\|_2} \cdot (\mathbf{q}_i - f(\mathbf{q}_i')) - u(\mathbf{q}_i)\right]^2 \tag{6}$$

In Fig. 5(c), $\frac{\nabla u(\mathbf{q}_3)}{\|\nabla u(\mathbf{q}_3)\|_2} \cdot (\mathbf{q}_3 - f(\mathbf{q}_3')) = \frac{\nabla u(\mathbf{q}_3)}{\|\nabla u(\mathbf{q}_3)\|_2} \cdot (\mathbf{q}_3 - \mathbf{t}_2)$ is the distance between $\mathbf{q}_3$ and $\mathbf{t}_2$ projected along the gradient direction $\frac{\nabla u(\mathbf{q}_3)}{\|\nabla u(\mathbf{q}_3)\|_2}$. By minimizing $[\frac{\nabla u(\mathbf{q}_3)}{\|\nabla u(\mathbf{q}_3)\|_2} \cdot (\mathbf{q}_3 - f(\mathbf{q}_3')) - u(\mathbf{q}_3)]^2$, we guide UltraBoneUDF to pull $\mathbf{q}_3$ toward the local bone surface.

**Global shape regularization.** $\mathcal{L}_{\text{tangent}}$ helps to pull the point toward the surface, but only using local geometric information. To facilitate faster global convergence in the beginning of the training, we introduce a regularization loss term for sparsely sampled grid anchor points. Specifically, as shown in Fig. 6, we sample $G$ anchor points $\rho_i$ uniformly within the bounding box of the preprocessed point cloud $\mathcal{P}$. These anchor points provide additional supervision for UltraBoneUDF to learn the global shape with the following loss term:

$$\mathcal{L}_{\text{GS}} = \frac{1}{G} \sum_{i\in[1,G]} [u(\rho_i) - \|\rho_i - f(\rho_i)\|_2]^2 \tag{7}$$

The final loss function of UltraBoneUDF is a weighted combination of $\mathcal{L}_{\text{tangent}}$ and $\mathcal{L}_{\text{GS}}$:

$$\mathcal{L}_{\text{UltraBoneUDF}} = \mathcal{L}_{\text{tangent}} + \lambda_{\text{GS}} \cdot \mathcal{L}_{\text{GS}} \tag{8}$$

Our model is trained in a self-supervised, per-instance manner, meaning each trained model encodes a single bone surface and does not require pretraining on other instances. For the 2D example in Fig. 5, the model trained with our proposed loss function consistently produces the correct UDF, as shown in Fig. 5(c).

### 3.3.3. Mesh extraction

After training the UDF predictor, direct surface reconstruction with algorithms such as Marching Cubes is infeasible because UDFs lack sign flips (Lorensen and Cline, 1998). In principle, the target surface corresponds to the zero-level set of the trained UltraBoneUDF. However, previous work (Zhang et al., 2023) has shown that neural UDFs tend to

**Table 1**
Quantitative results on the UltraBones100k dataset. (a) Results in terms of CD. (b) Results in terms of HD95. The best performing method is indicated in bold.

(a) Bi-directional CD between reconstruction and the CT bone model (mm).

| Methods | Specimen index | | | | | | | | | | | | | | |
|---|---|---|---|---|---|---|---|---|---|---|---|---|---|---|---|
| | #1 | #2 | #3 | #4 | #5 | #6 | #7 | #8 | #9 | #10 | #11 | #12 | #13 | #14 | mean |
| Poisson (Kazhdan et al., 2006) | 1.93 | **1.48** | 5.16 | 4.66 | 4.18 | 4.24 | 1.53 | **1.19** | 2.87 | 3.70 | 4.50 | 4.56 | 3.61 | 3.02 | 3.33 |
| FUNSR (Chen et al., 2024b) | 2.22 | 2.49 | 2.58 | 2.63 | 2.37 | 2.40 | 2.67 | 2.27 | 2.29 | 2.33 | 2.82 | 2.62 | 3.10 | 2.39 | 2.51 |
| CarotidUDF (Chen et al., 2023) | 3.13 | 3.58 | 3.32 | 3.46 | 3.26 | 3.21 | 3.45 | 3.29 | 3.19 | 3.35 | 3.79 | 3.60 | 4.06 | 3.18 | 3.42 |
| CAP-UDF (Zhou et al., 2024) | 1.60 | 3.27 | **1.33** | 2.44 | 2.97 | 2.71 | **1.30** | 4.08 | **1.23** | 2.80 | 1.96 | **1.72** | **1.76** | **0.99** | 2.15 |
| Ours | **1.08** | 1.56 | 1.64 | **1.55** | **1.45** | **1.42** | 1.44 | 1.33 | 1.53 | **1.72** | **1.70** | 2.16 | 2.35 | 1.47 | **1.60** |

(b) Bi-directional HD95 between reconstruction and the CT bone model (mm).

| Methods | Specimen index | | | | | | | | | | | | | | |
|---|---|---|---|---|---|---|---|---|---|---|---|---|---|---|---|
| | #1 | #2 | #3 | #4 | #5 | #6 | #7 | #8 | #9 | #10 | #11 | #12 | #13 | #14 | mean |
| Poisson (Kazhdan et al., 2006) | 9.01 | **5.51** | 27.76 | 26.47 | 22.38 | 25.83 | 5.88 | **4.14** | 15.71 | 20.06 | 26.28 | 27.71 | 18.98 | 18.69 | 18.17 |
| FUNSR (Chen et al., 2024b) | 6.13 | 6.92 | 7.40 | 7.01 | 6.52 | 6.54 | 8.07 | 6.33 | 6.32 | 6.57 | 8.69 | 8.15 | 10.61 | 6.49 | 7.27 |
| CarotidUDF (Chen et al., 2023) | 6.86 | 8.01 | 8.03 | 9.16 | 7.37 | 7.11 | 7.53 | 6.86 | 7.07 | 7.40 | 9.60 | 8.61 | 10.83 | 7.09 | 7.97 |
| CAP-UDF (Zhou et al., 2024) | 9.76 | 6.29 | **4.06** | 10.28 | 11.40 | 10.99 | **3.24** | 8.17 | **2.86** | 9.96 | 6.97 | **3.60** | **5.14** | **2.84** | 6.82 |
| Ours | **4.66** | 7.10 | 5.21 | **6.40** | **4.66** | **5.42** | 5.61 | 5.38 | 6.24 | **6.14** | **6.24** | 8.69 | 9.10 | 6.40 | **6.23** |

be noisy and imperfect, particularly in narrow regions near the surface and the cut locus. As described in Section 3.3.2, DualMesh-UDF takes these factors into account for mesh extraction. We employed DualMesh-UDF (Zhang et al., 2023) directly to extract the target surface from the trained UltraBoneUDF. For further details of DualMesh-UDF, we refer to Zhang et al. (2023).

## 4. Experiments

In the following sections, we first present the datasets used for benchmarking in Section 4.1, followed by a description of our model's implementation details in Section 4.2. Next, we outline the baseline methods used for comparison in Section 4.3. Finally, we discuss the evaluation metrics in Section 4.4 and the statistical evaluation in Section 4.5. Additional experiments such as ablation studies are presented in Section 4.6.

### 4.1. Datasets

For evaluation of models' performance and generalizability, we use 4 different datasets in our experiments.

- UltraBones100k (Wu et al., 2025). This open-source dataset comprises ex-vivo ultrasound scans from 14 human cadaveric specimens. Multiple scans were acquired from each specimen, encompassing the tibia, fibula, and foot bones, resulting in over 100k ultrasound images. Corresponding CT-based bone models and frame-wise tracking data are available for each specimen. Additionally, a pretrained model for segmenting bone anatomy in 2D ultrasound images is provided. During experiments, we directly applied their pretrained model to obtain bone segmentations from the ultrasound images. A raw point cloud is constructed using the segmentation masks and corresponding tracking data, which is further processed following the steps described in Section 3.1, including downsampling and query point sampling. The full CT-derived bone models are used as ground truth for evaluating the 3D reconstruction accuracy. UltraBoneUDF and other online self-supervised methods are trained separately for each scan across all specimens.
- OpenBoneCT. As ultrasound imaging captures only partial bone surfaces compared to CT, the CT-derived bone models of the UltraBones100k dataset, although spatially aligned, do not correspond to the bone structures visible in the ultrasound images. This limits the feasibility of bi-directional evaluation. To address this, we generate open CT-derived ground truth bone models by truncating the ultrasound-aligned CT-derived bone models of UltraBones100k with Blender (Blender Institute, Amsterdam, Netherlands). An example of the resulting truncated surface is shown in Fig. 2. During the experiments, raw point clouds are uniformly sampled from the CT-derived bone models and further processed following the steps outlined in Section 3.1. UltraBoneUDF and other online self-supervised methods are trained separately for each truncated bone surface across all specimens. Unlike UltraBones100k, the OpenBoneCT dataset serves as a benchmark dataset representing ideal conditions without external error sources, such as bone segmentation and tracking inaccuracies.
- ClosedBoneCT. To assess our model's generalizability for closed bone surfaces, we directly sample point clouds from the closed CT-derived bone models of the UltraBones100k dataset for bone surface reconstruction, which are further processed following the steps described in Section 3.1.
- Prostate. The open-source Prostate dataset contains transrectal ultrasound volumes annotated with labels for lesions, zonal structures, water-filled cysts, and other structures (Baum et al., 2023). This dataset has already been used to evaluate the SOTA SDF-based method FUNSR (Chen et al., 2024b). We applied it to all specimens in the validation split of the Prostate dataset (case #65 to #72). Specifically, points are uniformly sampled from the 3D ultrasound volume labels and provided as input to all models for label reconstruction. As this dataset is not directly relevant to CAOS, we use it primarily to assess the generalizability of UltraBoneUDF beyond CAOS applications. Hence, we report the results in the appendix.

### 4.2. Implementation details

Following the convention of previous work (Chen et al., 2024b; Ma et al., 2020), we set the size of the downsampled point cloud $\mathcal{P}$ to $|\mathcal{P}| = N = 40{,}000$. The number of query points per point $\mathbf{p}_i$ is set to $M = 20$, resulting in a final point cloud size of $L = M \cdot N = 800{,}000$. For the query point sampling strategy $\mathcal{N}(0, \delta)$, we define $\delta$ as the distance between $\mathbf{p}_i$ and its 50th nearest neighbor in $\mathcal{P}$. We set the batch size to $K = 5000$, the number of grid anchor points to $G = 1000$, and the weight of the regularization loss term to $\lambda_{GS} = 0.01$. UltraBoneUDF is trained for 30,000 iterations using the Adam optimizer with a learning rate of 0.001 and a momentum of 0.9. For mesh extraction, we directly apply the official implementation of DualMesh-UDF (Zhang et al., 2023) (depth=8, representing resolution $2^8 = 256$).[2] Following previous work (Chen et al., 2024b; Ma et al.,

[2] https://github.com/cong-yi/DualMesh-UDF

2020), we set the surface reconstruction resolution to 256 × 256 × 256 voxels. These parameters use the same values across different datasets and instances. UltraBoneUDF is implemented in PyTorch v2.7 with CUDA v12.8 on a system equipped with an NVIDIA H100 GPU. To ensure reproducibility, the code has been made publicly available.[3]

### 4.3. Baseline methods

- Poisson: Poisson surface reconstruction, a traditional method that estimates surfaces by solving a spatial Poisson equation over oriented point clouds, is commonly used for reconstructing surfaces from point clouds (Kazhdan et al., 2006). In our experiments, we employ the implementation provided in Open3D v0.18.0(Zhou et al., 2018), with a depth value of 9.
- FUNSR: FUNSR is a SOTA SDF-based method for anatomy surface reconstruction from 3D freehand ultrasound data scanned along a single direction (Chen et al., 2024b). In our experiments, we use the authors' official implementation.[4]
- CarotidUDF: CarotidUDF represents a SOTA UDF-based method for anatomy surface reconstruction from 3D freehand ultrasound data (Chen et al., 2023). Since the original code for CarotidUDF is not publicly available, we reimplemented the method in PyTorch following the details provided in the manuscript (Chen et al., 2023).
- CAP-UDF: CAP-UDF is a SOTA self-supervised network for learning neural UDFs from raw point clouds of general objects for surface reconstruction, with strong performance on synthetic datasets such as ShapeNet (Chang et al., 2015). It learns a consistency-aware UDF by iteratively moving query points toward the surface under a field-consistency constraint. Using the gradient directions, the learned UDF is converted into a pseudo-SDF, after which a variant of Marching Cubes is applied to extract the surface (Zhou et al., 2024). In our experiments, we use the authors' official implementation with default parameters (Zhou et al., 2024).[5]

### 4.4. Evaluation metrics

In all experiments, we compute the bi-directional CD and 95% Hausdorff Distance (HD95) between the US-reconstructed surface and the ground-truth CT-derived bone model. As it is infeasible to obtain exact ground-truth bone regions that are actually captured by the ultrasound data in the UltraBones100k dataset, we estimate a pseudo ground truth using distance thresholding between the ultrasound point cloud and the CT bone model ($\leq 2$ mm). The metrics are then computed based on these filtered meshes.

### 4.5. Statistical evaluation

To assess the significance of performance differences between UltraBoneUDF and the four competing methods, we use paired Wilcoxon signed-rank tests at a significance level of $\alpha = 0.05$, applying a Bonferroni correction to the resulting $p$-values ($p_{\text{corr}} = 4\,p_{\text{raw}}$). The tests are applied to all datasets.

[3] https://github.com/luohwu/UltraBoneUDF
[4] https://github.com/chenhbo/FUNSR
[5] https://github.com/junshengzhou/CAP-UDF

### 4.6. Additional experiments

**Hyperparameters.** The main hyperparameters of our method are the network depth and the loss weight $\lambda_{\text{GS}}$ of the global shape regularization term. To assess their influence, we evaluate models with depths in the range [2,10] and $\lambda_{\text{GS}}$ values in [0.0, 0.001, 0.01, 0.1, 1.0].

**Other regularization methods.** Our loss function is essentially a combination of dynamic nearest neighbor constraint and a gradient-based regularization. Other regularization methods have been applied in previous work to enhance the UDF training (Atzmon and Lipman, 2020; Gropp et al., 2020; Yariv et al., 2021). For comparison, we evaluated the widely used point-on-surface regularization and the Eikonal gradient regularization. The point-on-surface regularization $\mathcal{L}_{\text{OS}}$ (Eq. (9)) enforces the network to output zero UDF values for input points $\mathcal{P}$ (Atzmon and Lipman, 2020; Gropp et al., 2020). The Eikonal regularization $\mathcal{L}_{\text{Eik}}$ (Eq. (10)) enforces the network to produce UDF with unit-gradient. This makes the implicit field behave like a true distance function, improving geometric consistency and stabilizing surface extraction (Gropp et al., 2020; Yariv et al., 2021). We evaluate $\mathcal{L}_{\text{pull}} + \lambda_{\text{OS}} \cdot \mathcal{L}_{\text{OS}} + \lambda_{\text{Eik}} \cdot \mathcal{L}_{\text{Eik}}$ on the real-world UltraBones100k dataset with $\lambda_{\text{OS}}$ and $\lambda_{\text{Eik}}$ both in the range [0.1, 1, 10].

$$\mathcal{L}_{\text{OS}} = \frac{1}{|\mathcal{P}|} \sum_{\mathbf{p} \in \mathcal{P}} u(\mathbf{p})^2. \tag{9}$$

$$\mathcal{L}_{\text{Eik}} = \frac{1}{|\mathcal{P}| + |\mathcal{Q}|} \Big[ \sum_{\mathbf{p} \in \mathcal{P}} (\|\nabla u(\mathbf{p})\| - 1)^2 + \sum_{\mathbf{q} \in \mathcal{Q}} (\|\nabla u(\mathbf{q})\| - 1)^2 \Big] \tag{10}$$

**Robustness to noise.** Prior work (Wu et al., 2025; Pandey et al., 2020) shows that segmentation masks on target bones often contain false positives (causing false reconstructions) and false negatives (causing spatial gaps), which are already present in the UltraBones100k dataset. To further systematically assess the impact of such errors, we inject Gaussian noise $\mathcal{N}(0, \sigma^2)$ into points in the OpenBoneCT dataset with $\sigma \in [0.5, 1.0, 1.5, 2.0]$.

## 5. Results

In this section, we present both quantitative and qualitative results across all three datasets, followed by the results of the ablation experiments.

**UltraBones100k.** The mean CD and HD95 from the reconstruction to the CT-derived bone model for each specimen are shown in Table 1. UltraBoneUDF outperforms all methods with the lowest mean CD of 0.96 mm (28.9% improvement compared to SOTA) and HD95 of 2.56 mm (20.0% improvement compared to SOTA). The paired Wilcoxon signed-rank tests confirm that UltraBoneUDF significantly outperforms competing methods on both CD and HD95 ($p_{\text{corr}} < 0.001$). Example visualizations of reconstructed bone surfaces are shown in Fig. 7. The mean runtime is around 119 s on an NVIDIA H100 GPU.

**OpenBoneCT.** The mean bi-directional CD and HD95 between the reconstruction and the open CT-derived bone model for each specimen are reported in Table 2. UltraBoneUDF outperforms the other methods, achieving the lowest mean CD of 0.21 mm (40.0% improvement compared to SOTA) and HD95 of 0.90 mm (50.8% improvement compared to SOTA), with $p_{\text{corr}} < 0.05$. Example visualizations of reconstructed bone surfaces are shown in Fig. 8. Note that the surfaces reconstructed by FUNSR still exhibit holes. In contrast, UltraBoneUDF successfully reconstructs the target bone surfaces in all cases.

**ClosedBoneCT.** Similarly, the mean bi-directional CD and HD95 results for the ClosedBoneCT dataset are provided in Table 3. UltraBoneUDF shows the best performance with the lowest mean CD of 0.18 mm (63.3% improvement compared to SOTA) and HD95 of 0.44 mm (31.3% improvement compared to SOTA). The significance of the improvements is confirmed by the paired Wilcoxon signed-rank tests ($p_{\text{corr}} < 0.001$). Example visualizations of the reconstructed bone surfaces are shown in Fig. 9. Different from the results on OpenBoneCT,

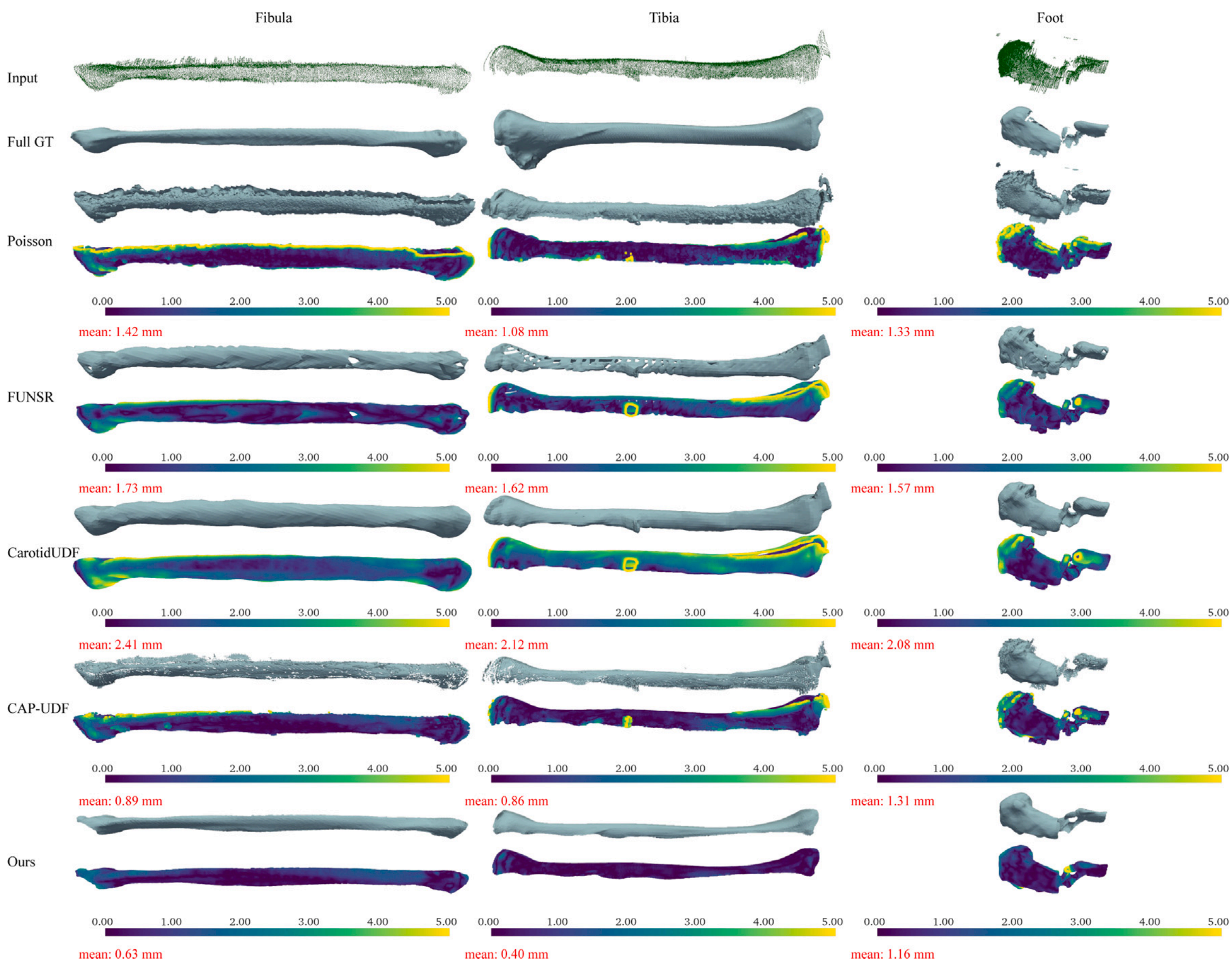

**Fig. 7.** Example visualizations of reconstructed bone surfaces on UltraBones100k. The colors encode the distance from the reconstructed mesh to the ground truth mesh, with darker colors indicating smaller values. For improved visualization, the displayed colors are limited to values in the range of 0–5 mm. Mean values are reported below the left edge of each color bar.

**Table 2**
Quantitative results on the OpenBoneCT dataset. (a) Results in terms of CD. (b) Results in terms of HD95. The best performing method is indicated in bold.

(a) Bi-directional CD between reconstruction and the CT-derived bone model (mm).

| Methods | Specimen index | | | | | | | | | | | | | | |
|---|---|---|---|---|---|---|---|---|---|---|---|---|---|---|---|
| | #1 | #2 | #3 | #4 | #5 | #6 | #7 | #8 | #9 | #10 | #11 | #12 | #13 | #14 | mean |
| Poisson (Kazhdan et al., 2006) | 0.51 | 0.69 | 0.64 | 0.81 | 0.64 | 0.95 | 0.89 | 0.66 | 0.85 | 0.57 | 0.74 | 0.92 | 0.86 | 0.72 | 0.75 |
| FUNSR (Chen et al., 2024b) | 1.22 | 1.42 | 1.22 | 1.35 | 1.36 | 1.16 | 1.50 | 1.37 | 1.16 | 1.32 | 1.26 | 1.39 | 1.15 | 1.21 | 1.29 |
| CarotidUDF (Chen et al., 2023) | 1.05 | 1.07 | 1.16 | 1.25 | 1.12 | 0.98 | 1.24 | 0.99 | 0.94 | 1.05 | 1.03 | 1.09 | 0.99 | 1.03 | 1.07 |
| CAP-UDF (Zhou et al., 2024) | **0.20** | 0.26 | 0.26 | 0.53 | 0.22 | 0.24 | 0.25 | 0.56 | 0.32 | 0.55 | 0.26 | 0.28 | 0.49 | 0.51 | 0.35 |
| Ours | 0.21 | **0.19** | **0.19** | **0.24** | **0.20** | **0.17** | **0.21** | **0.29** | **0.20** | **0.20** | **0.26** | **0.18** | **0.24** | **0.19** | **0.21** |

(b) Bi-directional HD95 between reconstruction and the CT-derived bone model (mm).

| Methods | Specimen index | | | | | | | | | | | | | | |
|---|---|---|---|---|---|---|---|---|---|---|---|---|---|---|---|
| | #1 | #2 | #3 | #4 | #5 | #6 | #7 | #8 | #9 | #10 | #11 | #12 | #13 | #14 | mean |
| Poisson (Kazhdan et al., 2006) | 2.95 | 3.99 | 3.65 | 4.38 | 3.74 | 5.53 | 5.02 | 3.62 | 5.02 | 3.20 | 4.28 | 5.26 | 5.00 | 4.08 | 4.27 |
| FUNSR (Chen et al., 2024b) | 2.74 | 3.07 | 2.92 | 3.11 | 2.93 | 2.60 | 3.91 | 3.65 | 2.71 | 2.76 | 2.75 | 2.95 | 2.55 | 2.66 | 2.95 |
| CarotidUDF (Chen et al., 2023) | 3.23 | 3.03 | 4.12 | 3.92 | 3.12 | 2.97 | 3.70 | 2.63 | 2.52 | 2.92 | 2.76 | 3.27 | 2.83 | 3.26 | 3.16 |
| CAP-UDF (Zhou et al., 2024) | **0.51** | 0.86 | 0.79 | 2.95 | 0.84 | 1.07 | 0.88 | 3.97 | 1.69 | 3.13 | **1.05** | 1.41 | 3.11 | 3.39 | 1.83 |
| Ours | 0.93 | **0.76** | **0.64** | **0.97** | **0.61** | **0.75** | **0.84** | **1.32** | **0.90** | **0.86** | 1.22 | **0.77** | **1.20** | **0.78** | **0.90** |

**Table 3**
Quantitative results on the ClosedBoneCT dataset. (a) Results in terms of CD. (b) Results in terms of HD95. The best performing method is indicated in bold.

(a) Bi-directional CD between reconstruction and the CT-derived bone model (mm).

| Methods | Specimen index | | | | | | | | | | | | | | |
|---|---|---|---|---|---|---|---|---|---|---|---|---|---|---|---|
| | #1 | #2 | #3 | #4 | #5 | #6 | #7 | #8 | #9 | #10 | #11 | #12 | #13 | #14 | mean |
| Poisson (Kazhdan et al., 2006) | 0.48 | 0.43 | 0.33 | 0.48 | 0.47 | 0.60 | 0.52 | 0.38 | 0.57 | 0.36 | 0.54 | 0.53 | 0.55 | 0.37 | 0.47 |
| FUNSR (Chen et al., 2024b) | 0.80 | 0.81 | 0.58 | 0.83 | 0.82 | 0.68 | 0.81 | 0.69 | 0.72 | 0.76 | 0.81 | 0.78 | 0.72 | 0.73 | 0.75 |
| CarotidUDF (Chen et al., 2023) | 1.06 | 1.04 | 0.79 | 1.12 | 1.03 | 0.87 | 1.06 | 0.90 | 0.91 | 0.97 | 0.97 | 0.96 | 0.92 | 0.95 | 0.97 |
| CAP-UDF (Zhou et al., 2024) | 0.50 | 0.50 | 0.45 | 0.52 | 0.51 | 0.46 | 0.51 | 0.48 | 0.47 | 0.50 | 0.51 | 0.47 | 0.47 | 0.48 | 0.49 |
| Ours | **0.17** | **0.17** | **0.08** | **0.21** | **0.16** | **0.16** | **0.27** | **0.20** | **0.13** | **0.22** | **0.21** | **0.12** | **0.15** | **0.20** | **0.18** |

(b) Bi-directional HD95 between reconstruction and the CT-derived bone model (mm).

| Methods | Specimen index | | | | | | | | | | | | | | |
|---|---|---|---|---|---|---|---|---|---|---|---|---|---|---|---|
| | #1 | #2 | #3 | #4 | #5 | #6 | #7 | #8 | #9 | #10 | #11 | #12 | #13 | #14 | mean |
| Poisson (Kazhdan et al., 2006) | 2.48 | 2.36 | 1.95 | 2.57 | 2.50 | 3.52 | 2.69 | 2.00 | 2.93 | 1.88 | 2.64 | 3.01 | 2.95 | 1.96 | 2.53 |
| FUNSR (Chen et al., 2024b) | 1.73 | 1.71 | 1.17 | 1.66 | 1.73 | 1.39 | 1.60 | 1.41 | 1.45 | 1.64 | 1.99 | 1.61 | 1.49 | 1.50 | 1.58 |
| CarotidUDF (Chen et al., 2023) | 2.18 | 2.18 | 1.60 | 2.22 | 2.24 | 1.79 | 2.13 | 1.86 | 1.81 | 2.06 | 1.97 | 2.05 | 1.93 | 1.94 | 2.00 |
| CAP-UDF (Zhou et al., 2024) | 0.68 | 0.66 | 0.55 | 0.69 | 0.68 | 0.58 | **0.66** | 0.61 | 0.60 | 0.66 | **0.83** | 0.61 | 0.59 | 0.61 | 0.64 |
| Ours | **0.35** | **0.31** | **0.19** | **0.46** | **0.45** | **0.30** | 0.88 | **0.37** | **0.32** | **0.61** | 1.04 | **0.23** | **0.22** | **0.38** | **0.44** |

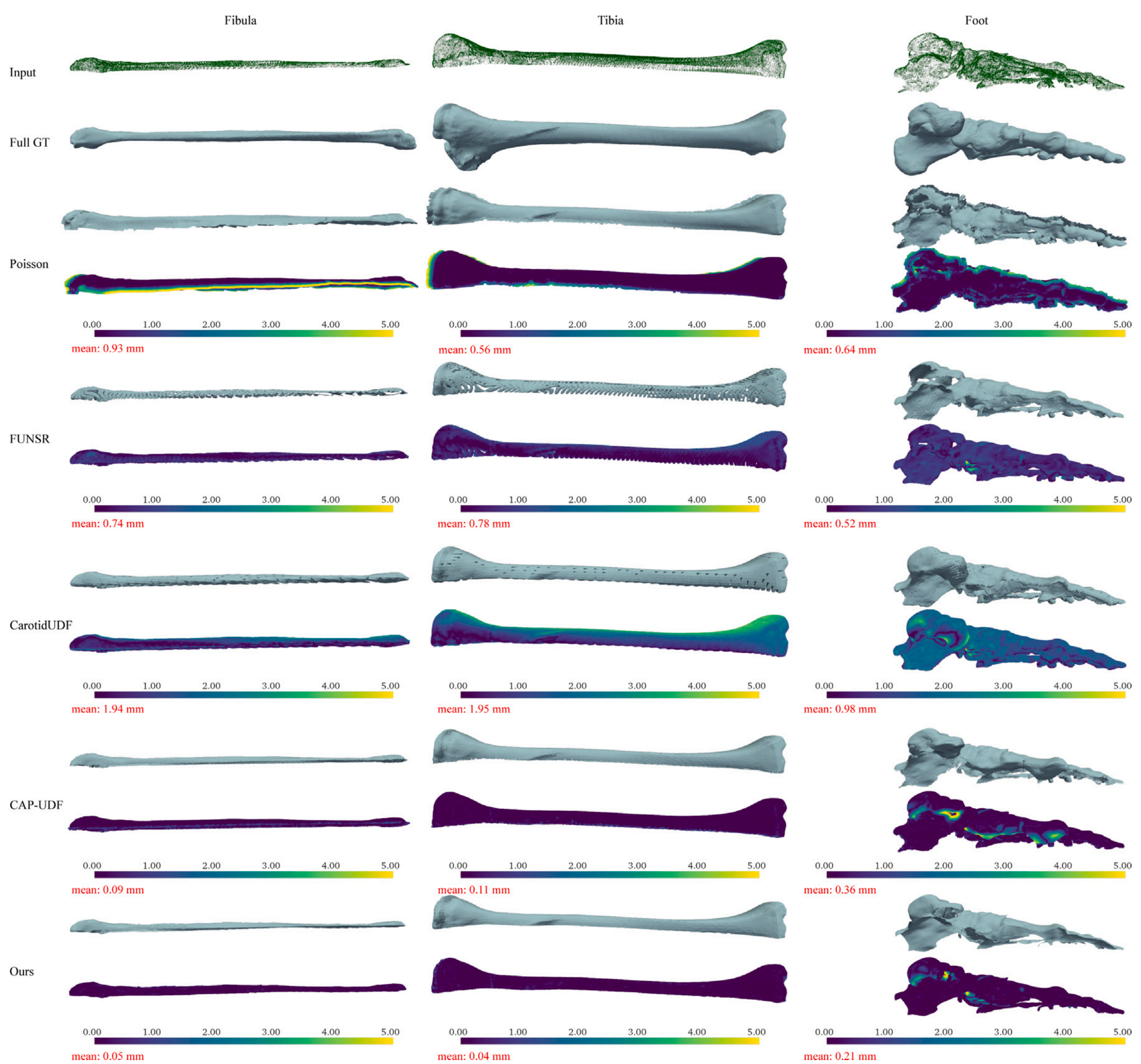

**Fig. 8.** Example visualizations of reconstructed bone surfaces on OpenBoneCT. The colors encode the distance from the reconstructed mesh to the ground truth mesh, with darker colors indicating smaller values. For improved visualization, the displayed colors are limited to values in the range of 0–5 mm. Mean values are reported below the left edge of each color bar.

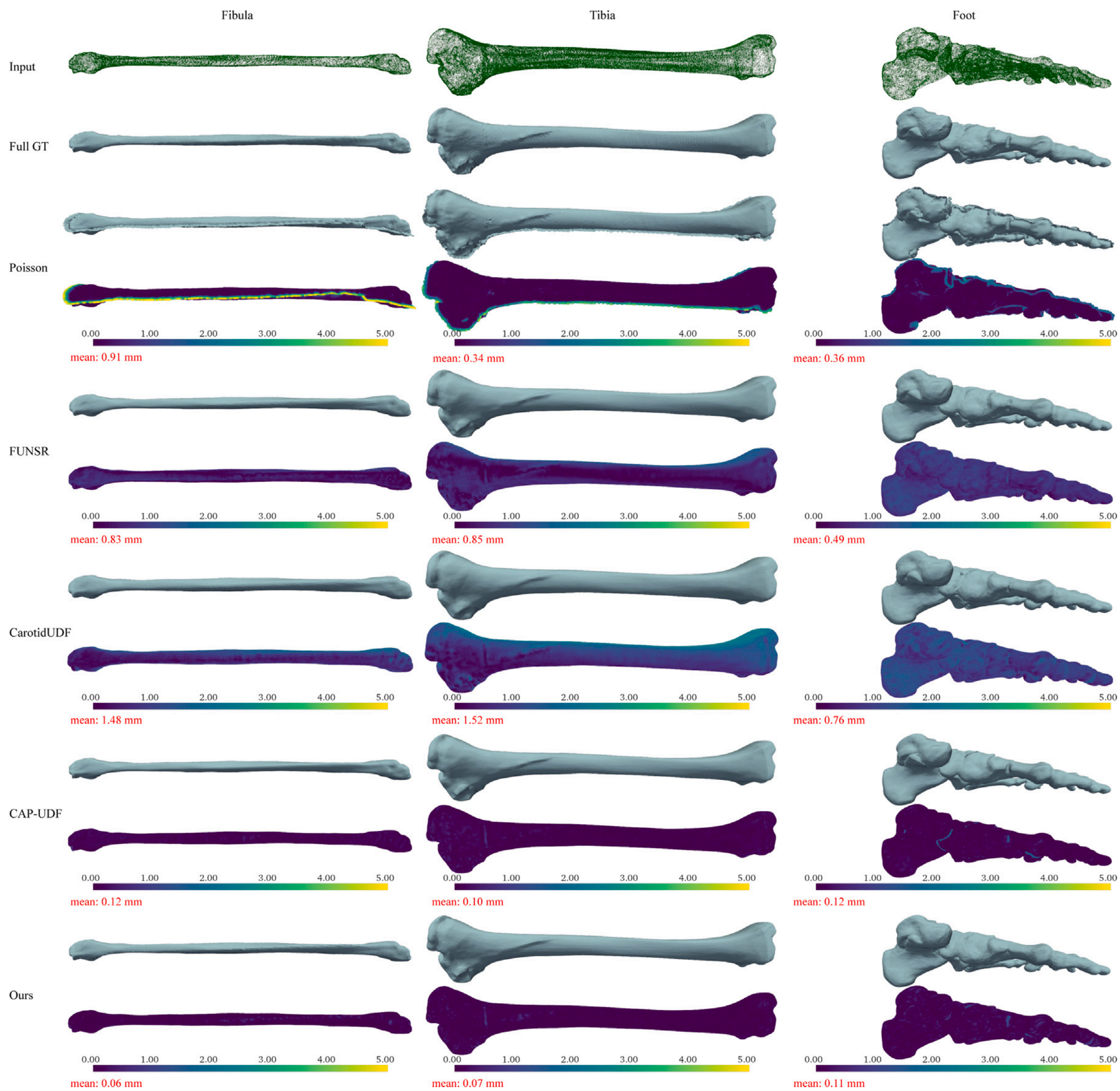

**Fig. 9.** Example visualizations of reconstructed bone surfaces on ClosedBoneCT. The colors encode the distance from the reconstructed mesh to the ground truth mesh, with darker colors indicating smaller values. For improved visualization, the displayed colors are limited to values in the range of 0–5 mm. Mean values are reported below the left edge of each color bar.

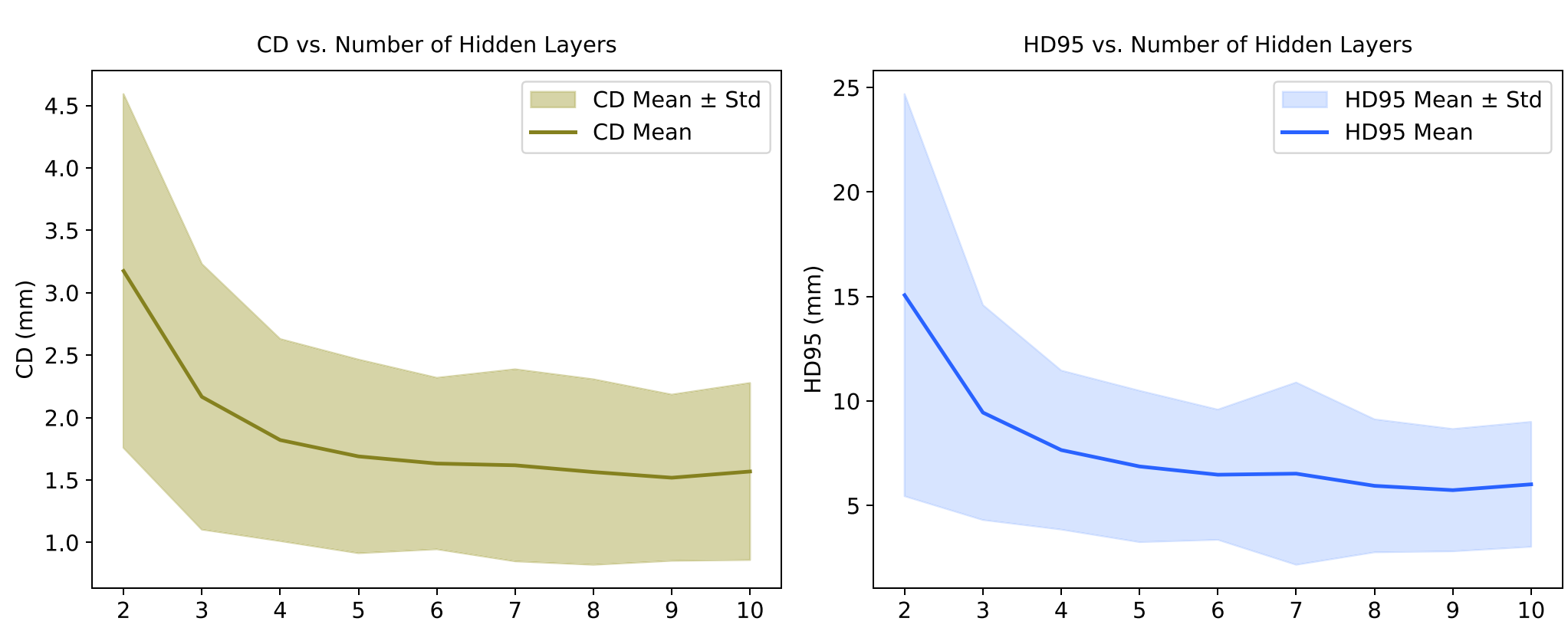

**Fig. 10.** Results of the empirical experiments on network depth.

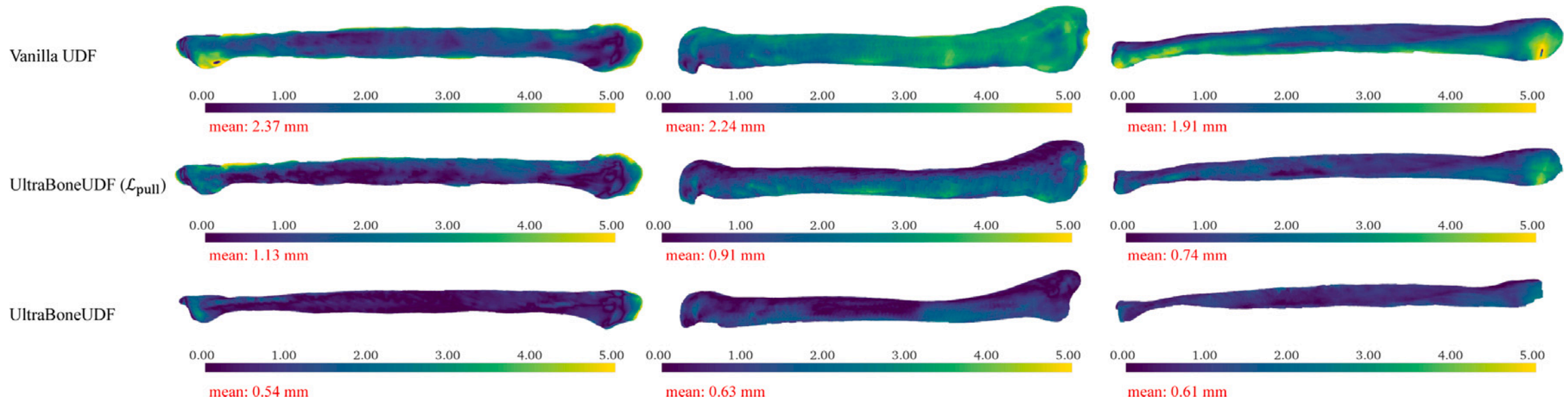

**Fig. 11.** Qualitative results of the ablation study on the UltraBones100k dataset. The colors encode the distance from the reconstructed mesh to the ground truth mesh, with darker colors indicating smaller values. For improved visualization, the displayed colors are limited to values in the range of 0–5 mm. Mean values are reported below the left edge of each color bar.

**Table 4**
Number of parameters and mean runtime of different methods.

| Methods | # Parameters | Mean runtime (s) |
|---|---|---|
| FUNSR (Chen et al., 2024b) | 595 453 | 183 |
| CarotidUDF (Chen et al., 2023) | 463 874 | 168 |
| CAP-UDF (Zhou et al., 2024) | 463 100 | 164 |
| Ours | 191 042 | 119 |

**Table 5**
Results of the empirical experiments on $\lambda_{\mathrm{GS}}$.

| $\lambda_{\mathrm{GS}}$ | CD (mm) | HD95 (mm) |
|---|---|---|
| 0.0 | 1.82 | 7.31 |
| 0.001 | 1.63 | 6.47 |
| 0.01 | 1.60 | 6.23 |
| 0.1 | 1.76 | 7.73 |
| 1.0 | 2.21 | 9.76 |

both SDF-based and UDF-based successfully reconstruct the target bone surfaces without artifacts.

**Model size.** The number of parameters and the mean runtime for different methods are reported in Table 4.

**Hyperparameters.** The influence of the network depth and the loss weight $\lambda_{\mathrm{GS}}$ are shown in Fig. 10 and Table 5 respectively. As shown by the results, the improvements are more significant when the network is shallower than 6 layers, and $\lambda_{\mathrm{GS}} = 0.01$ achieves the best performance.

**Other regularization methods.** The results of the point-on-surface regularization $\mathcal{L}_{\mathrm{OS}}$ and the Eikonal gradient regularization $\mathcal{L}_{\mathrm{Eik}}$ are reported in Table 6. As shown, the point-on-surface regularization $\mathcal{L}_{\mathrm{OS}}$ yields greater improvement than the gradient regularization $\mathcal{L}_{\mathrm{Eik}}$.

**Robustness to noise.** The results of the Gaussian noise experiments on the OpenBoneCT dataset are reported in Table 7. As $\sigma$ increases, both CD and HD95 consistently rise, reaching CD/HD95 of 1.48/6.39 mm at $\sigma = 2.0$.

**Ablation Study.** To assess the respective impact of the mesh extraction method and the loss function on reconstruction performance, we conduct an ablation study on the UltraBones100k dataset. We compare three models: the vanilla UDF, UltraBoneUDF($\mathcal{L}_{\mathrm{pull}}$), and UltraBoneUDF trained with our loss function (Eq. (8)). The vanilla UDF and UltraBoneUDF($\mathcal{L}_{\mathrm{pull}}$) differ only in the mesh extraction strategy, while UltraBoneUDF($\mathcal{L}_{\mathrm{pull}}$) and UltraBoneUDF differ only in the loss function. The qualitative results are shown in Fig. 11. The quantitative results are reported in Table 8. Specifically, compared to CarotidUDF, UltraBoneUDF ($\mathcal{L}_{\mathrm{pull}}$) improves the CD from 2.46 mm to 1.44 mm and the HD95 from 5.31 mm to 3.99 mm. UltraBoneUDF further improves the CD to 0.96 mm and the HD95 to 2.56 mm ($p_{\mathrm{corr}} < 0.001$).

## 6. Discussion

In this section, we provide a detailed discussion of our experimental results in terms of accuracy, noise robustness, runtime and limitations.

**Accuracy**. The UltraBones100k and OpenBoneCT datasets contain topologically open bone surfaces, whereas ClosedBoneCT provides topologically closed ones. SDF-based methods such as FUNSR assume that the target shape space can be partitioned into interior and exterior regions, an assumption that breaks down for open surfaces. The performance gaps of FUNSR between the OpenBoneCT and ClosedBoneCT datasets (e.g., Table 2 vs Table 3, Fig. 8 vs Fig. 9) highlight the limitation of SDFs, consistent with prior findings on UDFs for natural objects (Chibane and Pons-Moll, 2020; Zhou et al., 2024). In contrast, UDF-based methods perform similarly on both OpenBoneCT and ClosedBoneCT. Among UDF-based approaches, results across datasets and ablation studies (Table 8, Fig. 11) show that the loss functions and mesh extraction strategy substantially affect performance. CarotidUDF combines $\mathcal{L}_{\mathrm{pull}}$ with a simple UDF-to-SDF mesh extraction pipeline, leading to limited performance. CAP-UDF stabilizes UDF training using a consistency-aware loss with dynamic nearest-neighbor querying, achieving strong performance across datasets. Our method integrates dynamic nearest-neighbor querying with tangent-based optimization to train a UDF that is directly used for mesh extraction. These design choices align well with the challenges in CAOS ultrasound data, such as non-uniform point distributions and spatial gaps (Zhou et al., 2024; Zhang et al., 2023). The overall performance gains demonstrate the effectiveness of our designs. The ablation studies further support our hypothesis that aligning the training objective with the mesh extraction objective yields better initialization and improved optimization during mesh extraction. Essentially, the loss functions in both CAP-UDF and our method act as regularizers that stabilize UDF training. The results in Table 6 show that other regularization strategies are also beneficial, though with smaller performance gains, suggesting that more effective options may exist for our data.

**Robustness to noise.** The performance gaps observed across the real-world UltraBones100k dataset (Table 1) and the synthetic OpenBoneCT dataset (Table 2) highlight the impact of noise. The UltraBones100k dataset contains common real-world errors, including segmentation and tracking inaccuracies, reflecting challenges typical in clinical data. Across both datasets, Poisson shows a CD/HD95 difference of 2.57/13.9 mm, higher than other methods, underscoring the limitation of conventional approaches. In contrast, neural implicit representation-based methods are more robust to outliers and accurately reconstruct bone surfaces. Their ability to approximate continuous functions also yields smoother meshes, consistent with prior comparisons between traditional and neural implicit methods (Chen et al., 2024b). Nevertheless, neural reconstruction still suffers from noise, indicating that a denoising step could further improve performance. The noise can be broadly categorized into on-surface and off-surface errors. On-surface errors include false positives and false negatives near the target bone surfaces (Wu et al., 2025; Pandey et al., 2020). Table 7 shows that performance decreases as on-surface noise increases. Off-surface errors mainly arise from false positives caused by other anatomies (Wu et al., 2025; Pandey et al., 2020). While distinguishing on-surface true positives from off-surface false positives is difficult in individual 2D frames, a 3D point cloud aggregated from a continuous ultrasound sweep exhibits more consistent bone patterns than surrounding random noise, making noise suppression more feasible. This can be addressed using approaches such as statistical denoising or prior bone shape modeling.

**Table 6**
Results of experiments with $\lambda_{\text{Eik}}$ and $\lambda_{\text{OS}}$.

| | 1 | 2 | 3 | 4 | 5 | 6 | 7 | 8 | 9 | 10 | 11 | 12 | 13 | 14 | 15 | 16 |
|---|---|---|---|---|---|---|---|---|---|---|---|---|---|---|---|---|
| $\lambda_{\text{Eik}}$ | 0 | 0 | 0 | 0 | 0.1 | 0.1 | 0.1 | 0.1 | 1 | 1 | 1 | 1 | 10 | 10 | 10 | 10 |
| $\lambda_{\text{OS}}$ | 0 | 0.1 | 1 | 10 | 0 | 0.1 | 1 | 10 | 0 | 0.1 | 1 | 10 | 0 | 0.1 | 1 | 10 |
| CD | 2.62 | 2.33 | 2.34 | 2.47 | 2.68 | 2.85 | 2.91 | 2.86 | 3.52 | 3.43 | 3.73 | 3.51 | 4.68 | 4.31 | 4.79 | 6.52 |
| HD95 | 7.08 | 6.83 | 7.00 | 7.11 | 8.25 | 9.01 | 9.30 | 9.12 | 12.27 | 11.16 | 14.01 | 12.39 | 17.24 | 15.73 | 18.50 | 23.77 |

**Table 7**
Results of the experiments with Gaussian noise. $\sigma = 0$ means no noise is added.

| $\sigma$ | CD (mm) | HD95 (mm) |
|---|---|---|
| 0.0 | 0.21 | 0.90 |
| 0.5 | 0.49 | 2.26 |
| 1.0 | 0.87 | 4.53 |
| 1.5 | 1.06 | 4.97 |
| 2.0 | 1.48 | 6.39 |

**Runtime.** The number of parameters and the mean runtime of different methods are reported in Table 4. The hyperparameters of our network are selected based on empirical evaluations (e.g., Fig. 10 and Table 5), balancing runtime and performance. Although it matches or surpasses the performance of SOTA methods with fewer parameters and lower runtime, its runtime remains relatively long (≈119 s), limiting applicability in real-time intraoperative scenarios. This limitation mainly stems from the need for self-supervised methods to train from scratch on the fly.

**Limitation.** Our study has several limitations. Our method depends on accurate segmentation in B-mode ultrasound images, and segmentation errors can degrade its performance. A potential future direction is to incorporate a denoising module that leverages mutual information across multiple scans (Chen et al., 2024a). As demonstrated by RoCoSDF (Chen et al., 2024a), integrating 3D ultrasound data scanned from two different directions may further enhance the reconstruction accuracy and robustness (Chen et al., 2024a). In addition, its current runtime (≈119 s) is infeasible in real-time intraoperative scenarios. To shorten the training process, a potential direction is to incorporate prior bone shape information into the model, allowing it to avoid training from scratch (Park et al., 2019; Sitzmann et al., 2020). In our experiments, we evaluated the point-on-surface regularization $\mathcal{L}_{\text{OS}}$ and the Eikonal gradient regularization $\mathcal{L}_{\text{Eik}}$. More efficient regularization strategies for our task may be explored in future work.

## 7. Conclusion

UltraBoneUDF presents an effective solution for accurately reconstructing open bone surfaces from real-world 3D ultrasound data. Our results demonstrate the feasibility of high-fidelity open bone surface reconstruction from ultrasound for future CAOS applications. At the same time, challenges remain in optimizing computational efficiency and mitigating upstream error sources such as segmentation inaccuracies. Addressing these aspects will be essential for clinical translation, particularly in time-critical surgical workflows.

## CRediT authorship contribution statement

**Luohong Wu:** Writing – review & editing, Writing – original draft, Visualization, Validation, Software, Methodology, Formal analysis, Data curation, Conceptualization. **Matthias Seibold:** Writing – review & editing, Writing – original draft, Supervision, Project administration, Methodology, Investigation, Conceptualization. **Nicola A. Cavalcanti:** Validation, Resources, Data curation, Conceptualization. **Giuseppe Loggia:** Validation, Resources, Methodology, Conceptualization. **Lisa Reissner:** Validation, Resources, Investigation. **Bastian Sigrist:** Resources, Data curation. **Jonas Hein:** Resources, Methodology. **Lilian Calvet:** Writing – review & editing, Supervision. **Arnd Viehöfer:** Resources, Project administration, Methodology, Conceptualization. **Philipp Fürnstahl:** Writing – review & editing, Supervision, Resources, Project administration, Methodology, Funding acquisition, Conceptualization.

## Declaration of Generative AI and AI-assisted technologies in the writing process

During the preparation of this work the authors used ChatGPT in order to improve the readability and language of the manuscript. After using this tool, the authors reviewed and edited the content as needed and take full responsibility for the content of the published article.

## Declaration of competing interest

The authors declare that they have no known competing financial interests or personal relationships that could have appeared to influence the work reported in this paper.

## Acknowledgments

This research has been funded by the Innosuisse Flagship project PROFICIENCY No. PFFS-21-19. This work has also been supported by the OR-X, a Swiss national research infrastructure for translational surgery, and associated funding by the University of Zurich, Switzerland and University Hospital Balgrist.

## Appendix A. Supplementary data

Supplementary material related to this article can be found online at https://doi.org/10.1016/j.compmedimag.2025.102690.

## Data availability

We use open-source datasets. The source code is available at: https://github.com/luohwu/UltraBoneUDF.

**Table 8**
Quantitative results of the ablation study on the UltraBones100k dataset. (a) Results in terms of CD. (b) Results in terms of HD95.

(a) Bi-directional CD between reconstruction and the CT-derived bone model (mm).

| Methods | Specimen index | | | | | | | | | | | | | | |
|---|---|---|---|---|---|---|---|---|---|---|---|---|---|---|---|
| | #1 | #2 | #3 | #4 | #5 | #6 | #7 | #8 | #9 | #10 | #11 | #12 | #13 | #14 | mean |
| Vanilla UDF | 3.40 | 3.84 | 3.61 | 3.76 | 3.56 | 3.46 | 3.76 | 3.52 | 3.46 | 3.66 | 4.08 | 3.86 | 4.32 | 3.46 | 3.70 |
| UltraBoneUDF ($\mathcal{L}_{\text{pull}}$) | 1.63 | 1.88 | 2.95 | 2.78 | 2.31 | 2.54 | 2.67 | 2.09 | 2.74 | 3.12 | 3.43 | 3.33 | 2.94 | 2.31 | 2.62 |
| UltraBoneUDF | **1.08** | **1.56** | **1.64** | **1.55** | **1.45** | **1.42** | **1.44** | **1.33** | **1.53** | **1.72** | **1.70** | **2.16** | **2.35** | **1.47** | **1.60** |

(b) Bi-directional HD95 between reconstruction and the CT-derived bone model (mm).

| Methods | Specimen index | | | | | | | | | | | | | | |
|---|---|---|---|---|---|---|---|---|---|---|---|---|---|---|---|
| | #1 | #2 | #3 | #4 | #5 | #6 | #7 | #8 | #9 | #10 | #11 | #12 | #13 | #14 | mean |
| Vanilla UDF | 6.99 | 8.11 | 8.16 | 9.30 | 7.49 | 7.21 | 7.63 | 6.95 | 7.16 | 7.54 | 9.76 | 8.75 | 11.00 | 7.21 | 8.09 |
| UltraBoneUDF ($\mathcal{L}_{\text{pull}}$) | 5.49 | 7.76 | 7.61 | 8.46 | 7.34 | 8.89 | 8.69 | **4.90** | **5.45** | 6.90 | 8.11 | **7.60** | **5.48** | 6.44 | 7.08 |
| UltraBoneUDF | **4.66** | **7.10** | **5.21** | **6.40** | **4.66** | **5.42** | **5.61** | 5.38 | 6.24 | **6.14** | **6.24** | 8.69 | 9.10 | **6.40** | **6.23** |

## References

Atzmon, M., Lipman, Y., 2020. Sal: Sign agnostic learning of shapes from raw data. In: Proceedings of the IEEE/CVF Conference on Computer Vision and Pattern Recognition. pp. 2565–2574.

Baraza, N., Chapman, C., Zakani, S., Mulpuri, K., 2020. 3D-printed patient specific instrumentation in corrective osteotomy of the femur and pelvis: a review of the literature. 3D Printing in Med. 6, 1–7.

Baum, Z., Saeed, S., Min, Z., Hu, Y., Barratt, D., 2023. MR to ultrasound registration for prostate challenge-dataset. In: Medical Image Computing and Computer Assisted Intervention. MICCAI 2023, Zenodo, Zenodo.

Berger, M., Tagliasacchi, A., Seversky, L.M., Alliez, P., Guennebaud, G., Levine, J.A., Sharf, A., Silva, C.T., 2017. A survey of surface reconstruction from point clouds. In: Computer Graphics Forum, vol. 36, (1), Wiley Online Library, pp. 301–329.

Bhatnagar, B.L., Tiwari, G., Theobalt, C., Pons-Moll, G., 2019. Multi-garment net: Learning to dress 3D people from images. In: IEEE International Conference on Computer Vision. ICCV, IEEE.

Chang, A.X., Funkhouser, T., Guibas, L., Hanrahan, P., Huang, Q., Li, Z., Savarese, S., Savva, M., Song, S., Su, H., Jianxiong, X., Li, Y., Fisher, Y., 2015. Shapenet: An information-rich 3d model repository. arXiv preprint arXiv:1512.03012.

Chen, H., Gao, Y., Zhang, S., Wu, J., Ma, Y., Zheng, R., 2024a. RoCoSDF: Row-column scanned neural signed distance fields for freehand 3D ultrasound imaging shape reconstruction. In: International Conference on Medical Image Computing and Computer-Assisted Intervention. Springer, pp. 721–731.

Chen, H., Kumaralingam, L., Li, J., Punithakumar, K., Le, L.H., Zheng, R., 2023. Neural implicit representation for three-dimensional ultrasound carotid surface reconstruction using unsigned distance function. In: 2023 IEEE International Ultrasonics Symposium. IUS, IEEE, pp. 1–3.

Chen, H., Kumaralingam, L., Zhang, S., Song, S., Zhang, F., Zhang, H., Pham, T.-T., Punithakumar, K., Lou, E.H., Zhang, Y., Le, L.H., Zheng, R., 2024b. Neural implicit surface reconstruction of freehand 3D ultrasound volume with geometric constraints. Med. Image Anal. 98, 103305, URL https://www.sciencedirect.com/science/article/pii/S1361841524002305.

Chibane, J., Pons-Moll, G., 2020. Neural unsigned distance fields for implicit function learning. In: Advances in Neural Information Processing Systems, vol. 33, pp. 21638–21652.

Ciganovic, M., Ozdemir, F., Pean, F., Fuernstahl, P., Tanner, C., Goksel, O., 2018. Registration of 3D freehand ultrasound to a bone model for orthopedic procedures of the forearm. Int. J. Comput. Assist. Radiol. Surg. 13 (6), 827–836. http://dx.doi.org/10.1007/s11548-018-1756-0.

Evrard, R., Schubert, T., Paul, L., Docquier, P.-L., 2022. Quality of resection margin with patient specific instrument for bone tumor resection. J. Bone Oncol. 34, 100434.

Fadero, P.E., Shah, M., 2014. Three dimensional (3D) modelling and surgical planning in trauma and orthopaedics. Surg. 12 (6), 328–333.

Gebhardt, C., Göttling, L., Buchberger, L., Ziegler, C., Endres, F., Wuermeling, Q., Holzapfel, B.M., Wein, W., Wagner, F., Zettinig, O., 2023. Femur reconstruction in 3D ultrasound for orthopedic surgery planning. Int. J. Comput. Assist. Radiol. Surg. 18 (6), 1001–1008.

Gropp, A., Yariv, L., Haim, N., Atzmon, M., Lipman, Y., 2020. Implicit geometric regularization for learning shapes. arXiv preprint arXiv:2002.10099.

Guezou-Philippe, A., Dardenne, G., Letissier, H., Yvinou, A., Burdin, V., Stindel, E., Lefevre, C., 2023. Anterior pelvic plane estimation for total hip arthroplasty using a joint ultrasound and statistical shape model based approach. Med. Biol. Eng. Comput. 61 (1), 195–204.

Guillard, B., Stella, F., Fua, P., 2022. Meshudf: Fast and differentiable meshing of unsigned distance field networks. In: European Conference on Computer Vision. Springer, pp. 576–592.

Hohlmann, B., 2024. Ultrasound-Based 3D Reconstruction of the Knee Joint (Ph.D. thesis, Dissertation). Rheinisch-Westfälische Technische Hochschule Aachen, 2024.

Hohlmann, B., Broessner, P., Phlippen, L., Rohde, T., Radermacher, K., 2023. Knee bone models from ultrasound. IEEE Trans. Ultrason. Ferroelectr. Freq. Control 70 (9), 1054–1063. http://dx.doi.org/10.1109/TUFFC.2023.3286287.

Hohlmann, B., Broessner, P., Radermacher, K., 2024. Ultrasound-based 3D bone modelling in computer assisted orthopedic surgery - a review and future challenges. Comput. Assist. Surg. (Abingdon, England) 29, 2276055.

Kazhdan, M., Bolitho, M., Hoppe, H., 2006. Poisson surface reconstruction. In: Proceedings of the Fourth Eurographics Symposium on Geometry Processing, vol. 7, (4).

Li, R., Davoodi, A., Cai, Y., Niu, K., Borghesan, G., Cavalcanti, N., Massalimova, A., Carrillo, F., Laux, C.J., Farshad, M., Fürnstahl, P., Poorten, E.V., 2023. Robot-assisted ultrasound reconstruction for spine surgery: from bench-top to pre-clinical study. Int. J. Comput. Assist. Radiol. Surg. 18 (9), 1613–1623. http://dx.doi.org/10.1007/s11548-023-02932-z.

Lorensen, W.E., Cline, H.E., 1998. Marching cubes: A high resolution 3D surface construction algorithm. In: Seminal Graphics: Pioneering Efforts that Shaped the Field. pp. 347–353.

Ma, B., Han, Z., Liu, Y.-S., Zwicker, M., 2020. Neural-pull: Learning signed distance functions from point clouds by learning to pull space onto surfaces. URL https://arxiv.org/abs/2011.13495.

Mahfouz, M.R., Fatah, E.E.A., Johnson, J.M., Komistek, R.D., 2021. A novel approach to 3D bone creation in minutes: 3D ultrasound. Bone Jt. J. 103 (6 Supple A), 81–86.

Mavrogenis, A.F., Savvidou, O.D., Mimidis, G., Papanastasiou, J., Koulalis, D., Demertzis, N., Papagelopoulos, P.J., 2013. Computer-assisted navigation in orthopedic surgery. Orthopedics 36 (8), 631–642.

Mildenhall, B., Srinivasan, P.P., Tancik, M., Barron, J.T., Ramamoorthi, R., Ng, R., 2021. Nerf: Representing scenes as neural radiance fields for view synthesis. Commun. ACM 65 (1), 99–106.

Mozaffari, M.H., Lee, W.-S., 2017. Freehand 3-D ultrasound imaging: a systematic review. Ultrasound Med. Biol. 43 (10), 2099–2124.

Nguyen, D.V., Vo, Q.N., Le, L.H., Lou, E.H., 2015. Validation of 3D surface reconstruction of vertebrae and spinal column using 3D ultrasound data–A pilot study. Med. Eng. Phys. 37 (2), 239–244.

Pandey, P.U., Quader, N., Guy, P., Garbi, R., Hodgson, A.J., 2020. Ultrasound bone segmentation: A scoping review of techniques and validation practices. Ultrasound Med. Biol. URL https://api.semanticscholar.org/CorpusID:210924028.

Park, J.J., Florence, P., Straub, J., Newcombe, R., Lovegrove, S., 2019. Deepsdf: Learning continuous signed distance functions for shape representation. In: Proceedings of the IEEE/CVF Conference on Computer Vision and Pattern Recognition. pp. 165–174.

Schumann, S., Nolte, L.-P., Zheng, G., 2012. Determination of pelvic orientation from sparse ultrasound data for THA operated in the lateral position. Int. J. Med. Robot. Comput. Assist. Surg. 8 (1), 107–113.

Sitzmann, V., Chan, E., Tucker, R., Snavely, N., Wetzstein, G., 2020. Metasdf: Meta-learning signed distance functions. In: Advances in neural information processing systems, vol. 33, pp. 10136–10147.

So, T.Y., Lam, Y.-L., Mak, K.-L., 2010. Computer-assisted navigation in bone tumor surgery: seamless workflow model and evolution of technique. Clin. Orthop. Relat. Res.® 468, 2985–2991.

Sugano, N., 2003. Computer-assisted orthopedic surgery. J. Orthop. Sci. 8, 442–448.

Tetsworth, K., Block, S., Glatt, V., 2017. Putting 3D modelling and 3D printing into practice: virtual surgery and preoperative planning to reconstruct complex post-traumatic skeletal deformities and defects. Sicot-J 3, 16.

Venkatesh, R., Sharma, S., Ghosh, A., Jeni, L., Singh, M., 2020. Dude: Deep unsigned distance embeddings for hi-fidelity representation of complex 3d surfaces. arXiv preprint arXiv:2011.02570.

Wu, L., Cavalcanti, N.A., Seibold, M., Loggia, G., Reissner, L., Hein, J., Beeler, S., Viehöfer, A., Wirth, S., Calvet, L., Fürnstahl, P., 2025. UltraBones100k: A reliable automated labeling method and large-scale dataset for ultrasound-based bone surface extraction. Comput. Biol. Med. 194, 110435. http://dx.doi.org/10.1016/j.compbiomed.2025.110435, URL https://www.sciencedirect.com/science/article/pii/S0010482525007863.

Yariv, L., Gu, J., Kasten, Y., Lipman, Y., 2021. Volume rendering of neural implicit surfaces. In: Advances in neural information processing systems, vol. 34, pp. 4805–4815.

van der Zee, J.M., Fitski, M., van de Sande, M.A.J., Buser, M.A.D., Hiep, M.A.J., Terwisscha van Scheltinga, C.E.J., Hulsker, C.C.C., van den Bosch, C.H., van de Ven, C.P., van der Heijden, L., Bökkerink, G.M.J., Wijnen, M.H.W.A., Siepel, F.J., van der Steeg, A.F.W., 2023. Tracked ultrasound registration for intraoperative navigation during pediatric bone tumor resections with soft tissue components: a porcine cadaver study. Int. J. Comput. Assist. Radiol. Surg. 19 (2), 297–302. http://dx.doi.org/10.1007/s11548-023-03021-x.

Zhang, C., Lin, G., Yang, L., Li, X., Komura, T., Schaefer, S., Keyser, J., Wang, W., 2023. Surface extraction from neural unsigned distance fields. In: Proceedings of the IEEE/CVF International Conference on Computer Vision. pp. 22531–22540.

Zhou, J., Ma, B., Li, S., Liu, Y.-S., Fang, Y., Han, Z., 2024. CAP-UDF: Learning unsigned distance functions progressively from raw point clouds with consistency-aware field optimization. IEEE Trans. Pattern Anal. Mach. Intell.

Zhou, Q.-Y., Park, J., Koltun, V., 2018. Open3D: A modern library for 3D data processing. arXiv:1801.09847.